\documentclass[11pt,reqno]{amsart}
\RequirePackage[authoryear,comma]{natbib}
\RequirePackage[margin=1.2in]{geometry}
\RequirePackage{multirow}
\usepackage{pkgfile}
\usepackage{mathtools,subcaption}

\newcommand\independent{\protect\mathpalette{\protect\independenT}{\perp}}
\def\independenT#1#2{\mathrel{\rlap{$#1#2$}\mkern2mu{#1#2}}}

\newcommand*\diff{\mathop{}\!\mathrm{d}}

\newlength{\dhatheight}

\usepackage{amsaddr}
\allowdisplaybreaks

\usepackage{xr}
\usepackage{color}
\definecolor{qz}{RGB}{255,0,0}

\definecolor{blue}{rgb}{0,0,0}

\begin{document}

\title[Sensitivity Analysis for IPW]{Sensitivity Analysis
  for Inverse Probability Weighting Estimators via the Percentile
  Bootstrap}

\author{Qingyuan Zhao, Dylan S.\ Small and Bhaswar B.\ Bhattacharya}
\address{Department of Statistics, The Wharton School, University of
  Pennsylvania}

\email{\{qyzhao,dsmall,bhaswar\}@wharton.upenn.edu}
\date{\today}

\maketitle

\begin{abstract}
  To identify the estimand in missing data problems and observational
  studies, it is common to base the statistical estimation on the
  ``missing at random'' and ``no unmeasured confounder''
  assumptions. However, these assumptions are unverifiable using
  empirical data and pose serious threats to the validity of the
  qualitative conclusions of the statistical inference. A sensitivity
  analysis asks how the conclusions may change if the unverifiable
  assumptions are violated to a certain degree. In this paper we
  consider a marginal sensitivity model which is a natural extension
  of Rosenbaum's sensitivity model that is widely used for matched observational
  studies. We aim to construct confidence intervals based on inverse
  probability weighting estimators, such that asymptotically the intervals have
  at least nominal coverage of the estimand whenever the data
  generating distribution is in the collection of marginal sensitivity models.
  We use a percentile bootstrap and a generalized
  minimax/maximin inequality to transform this intractable
  problem to a linear fractional programming problem, which can be
  solved very efficiently. We illustrate our method using a real dataset
  to estimate the causal effect of fish consumption on blood mercury
  level.
\end{abstract}

\section{Introduction}
\label{sec:introduction}

A common task in statistics is to estimate the average treatment
effect in observational studies. In such problems, the estimand is
not identifiable from the observed data without further assumptions. The
most common identification assumption, namely the ``no unmeasured
confounder'' (NUC) assumption, asserts that the
confounding mechanism is completely
determined by some observed covariates. Based on this assumption, many
statistical procedures have been proposed and thoroughly studied in
the past decades, including propensity score matching
\citep{rosenbaum1983central}, inverse probability weighting
\citep{horvitz1952generalization}, and doubly robust and machine
learning estimators
\citep{robins1994estimation,van2006targeted,chernozhukov2017double}.


However, the underlying NUC assumption is not verifiable using
empirical data, posing serious threats to the usefulness of the
subsequent statistical inferences that crucially rely on this
assumption. A prominent example is the antioxidant vitamin
beta carotene. \citet{willett1990vitamin}, after reviewing
observational epidemiological data,  concluded that ``Available data
thus strongly support the hypothesis that dietary carotenoids reduce
the risk of lung cancer''. Quite unexpectedly, four years later a
large-scale randomized controlled trial  \citep{betacarotene1994effect} reached the opposite conclusion and
found ``a higher incidence of
lung cancer among the men who received beta carotene than among those
who did not (change in incidence, 18 percent; 95 percent confidence
interval, 3 to 36 percent)''. The most
probable reason for the disagreement between the observational studies and
the randomized trial is insufficient control of
confounding. \blue{People taking vitamin supplements tend to be
  healthier at baseline and have higher socioeconomic position, and
  the adjustment for social and environmental confounding may be
  insufficient in many observational studies
  \citep{lawlor2004those}. For the same reason, observational
  studies and randomized controlled trials often give different
  conclusions for other vitamin supplements
  \citep{lawlor2004those,rutter2007identifying}. We refer the reader to
  \citet[Section 6.4]{rutter2007identifying} for other more recent examples of
  observational studies with probably misleading causal claims due to
  unmeasured confounding.}

Criticism of confounding bias in observational studies dates at least back
to \citet{fisher1958lung} who suggested that the association between
smoking and lung cancer may be due to genetic predisposition. In
response to this, \citet{cornfield1959smoking} conducted the first
formal sensitivity analysis in an observational study. They concluded
that, in order to explain the apparent odds ratio of getting lung
cancer when there is no real causal effect, those with the genetic
predisposition (or any hypothetical unmeasured confounder) must be 9
times more prevalent in smokers than in non-smokers. Because a genetic predisposition was seen as unlikely to have such a strong effect, this strengthened the evidence that smoking had a causal harmful effect.

\citet{cornfield1959smoking}'s initial sensitivity
analysis only applies to binary outcomes and ignores sampling
variability. These limitations were later removed in a series of pioneering work by Rosenbaum and his coauthors \citep{rosenbaum1987sensitivity,gastwirth1998dual,rosenbaum2002covariance,rosenbaum2002observational}. Rosenbaum's
sensitivity model considers all possible violations of the NUC
assumption as long as the violation is less than some degree. Using a
matched cohort design, Rosenbaum attempts
to quantify the strength of the unmeasured confounders needed to not
reject the sharp null hypothesis of no treatment effect. However,
Rosenbaum's framework is only limited to matched observational studies
and usually requires effect homogeneity to construct confidence
intervals for the average treatment effect.

Besides matching, another widely used method in causal
inference is inverse probability weighting (IPW), where the
observations are weighted by the inverse of the probability of being
treated \citep{horvitz1952generalization}. IPW
estimators have good efficiency properties \citep{hirano2003efficient}
and can be augmented with outcome regression to become ``doubly
robust'' \citep{robins1994estimation}. There is much recent work
aiming to improve the efficiency of IPW-type estimators
by using tools developed in machine learning and high dimensional
statistics
\citep{van2011targeted,belloni2014inference,athey2016approximate,chernozhukov2017double}.
However, they all heavily rely on the NUC assumption. Robustness of
the IPW-type estimators to unmeasured confounding bias are usually studied
using pattern-mixture models \citep{birmingham2003pattern} or selection
models \citep{scharfstein1999adjusting}, in which the unmeasured
confounding is usually modeled parametrically. See
\citet{richardson2014nonparametric} for a recent overview and
\Cref{sec:related-work} for more references and discussion.


In this paper we propose a new framework for sensitivity analysis of
missing data and observational studies problems, which can be applied
to ``smooth'' estimators such as the inverse probability weighting (IPW)
estimator and the ``doubly robust'' augmented IPW estimators. We
consider a marginal sensitivity model introduced in \citet[Section
4.2]{tan2006distributional}, which is a natural modification of Rosenbaum's
model. Compared to existing sensitivity analyses of IPW estimators,
our sensitivity model is nonparametric in the sense that we do not
require the unmeasured confounder to affect the treatment and the
outcome in a parametric way (though the propensity score can be
modeled in a parametric way). This is appealing because we can never
observe the unmeasured confounder and thus cannot test any
parametric assumption. 

Our marginal sensitivity model measures the degree of violation of the
NUC assumption by the odds ratio between the conditional probability
of being treated given the measured confounders and conditional
probability of being treated given the measured confounders
and the outcome/potential outcome variable (see
\Cref{sec:sensitivity-models} for the precise definition). Given a
user-specified magnitude  for this odds-ratio, the goal is to obtain a
confidence interval of the estimand (the mean response vector in
missing data problems and the average treatment effect in
observational studies), with asymptotically at least $(1-\alpha)$ coverage
probability, for all data generating distributions which violate the
MAR assumption within this threshold (see Definition
\ref{defn:conf-int-sens-model}). \blue{Additionally, we would like to
  also report an interval of point estimates which is the range of
  possible point estimates under a sensitivity model. Our main
  contribution in this paper is to show such an interval of point
  estimates can be computed very efficiently for IPW estimators, and
  a valid confidence interval for sensitivity analysis can then be
  obtained by bootstrapping the interval of point estimates. The
  relationship to \citet{efron1979bootstrap}'s Bootstrap is explained
  below:}
\newcommand{\tR}[1]{\textcolor{Red}{\textbf{#1}}}
\begin{itemize}
\item Efron's bootstrap for a point-identified parameter: \\
  \begin{center} \vspace{-5pt}
    \begin{tabular}{ccc}
      & {\small \tR{\it Bootstrap}} & \\
      Point estimator & $\xRightarrow{\hspace*{2.5cm}}$ & Confidence
                                                          interval \\
    \end{tabular}
  \end{center}
\item Proposed percentile bootstrap procedure for sensitivity analysis
  (partially identified parameter):\\
  \begin{center} \vspace{-5pt}
    \begin{tabular}{ccc}
      {\small \tR{\it Optimization}} & {\small \tR{\it Percentile Bootstrap}}
      & {\small \tR{\it Minimax inequality}} \\
      Interval of point estimates & $\xRightarrow{\hspace*{2.5cm}}$ & Confidence interval \\
    \end{tabular}
  \end{center}
\end{itemize}


The obvious and direct approach to obtain confidence intervals in a
sensitivity analysis involves using the asymptotic normal distribution of the IPW estimators. However, the asymptotic sandwich variance estimators of the IPW estimates are quite complicated. Finding extrema
of the point estimates and variance estimates over a collection of
sensitivity models is generally computationally intractable. This
problem can be circumvented using the percentile bootstrap, reducing
the problem to solving many linear fractional programs. The confidence
interval our procedure constructs has the following desirable properties:

\begin{enumerate}

\item \blue{The interval covers the entire partially identified region of the estimand with asymptotic probability of the coverage at least $(1-\alpha)$, that is, it has asymptotically `strong nominal coverage' \citep[Definition 3]{vansteelandt2006ignorance}.} The proof involves establishing the limiting normal distribution of the bootstrapped IPW estimators and a generalized minimax/maximin inequality, which justifies the interchange of quantile and infimum/supremum (Theorem \ref{thm:valid-pb} and Corollary \ref{cor:valid-pb-obs}).

\item Maximizing/minimizing the IPW point estimate over the marginal
  sensitivity model is very efficient computationally. This is a
  linear fractional programming problem (ratio of two linear
  functions) which can be reduced to solving a linear program using
  the well-known Charnes-Cooper transformation (Section
  \ref{sec:line-fract-progr}). In fact, by local perturbations, it can
  be shown that the solution of the resulting linear program has the
  same/opposite order as the outcome vectors, using which the
  confidence interval can be computed in time linear in sample size, for every
  bootstrap resample (Proposition \ref{ppn:lfp}).

\end{enumerate}


Our method for constructing confidence intervals under the marginal sensitivity model, for the
IPW estimator for the mean response with missing data, is described in
\Cref{sec:new-framework}.  The extension to estimating the average
treatment effect in observational studies is discussed in
\Cref{sec:sens-analys-observ}. The percentile bootstrap approach and
the reduction to linear programming are very general and can be
extended to sensitivity analysis of other smooth estimators, such as
IPW estimates of the mean of the non-respondents and the average
treatment effect on the treated (\Cref{sec:mean-non-respondents}), the
augmented inverse probability weighting estimator
(\Cref{sec:augm-inverse-prob}), and inference for partially identified
parameters (\Cref{sec:part-ident-param}).  We review some related
sensitivity analysis methods and compare our framework with
Rosenbaum's sensitivity analysis in \Cref{sec:discussion}. In
\Cref{sec:numer-illustr} we evaluate the performance of our method
using some numerical examples, including a simulation study in
\Cref{sec:simulation-study} and a real data example in
\Cref{sec:real-data-example} where Rosenbaum's sensitivity
analysis is also applied. Theoretical proofs can be found in
\Cref{sec:proofs-main} and the supplementary file. The \texttt{R} code
for the our method and the real data example in
\Cref{sec:real-data-example} can be found at
\url{https://github.com/qingyuanzhao/bootsens}.

\section{The Missing Data Problem: Background and Notation}
\label{sec:background}

In this paper, we consider sensitivity analysis for the estimation of the mean response with missing data. The framework easily extends to estimation of the average treatment effect in observational studies (details given in \Cref{sec:sens-analys-observ}). In the missing data problem, we assume $(A_1,\bm X_1,Y_1)$,
$(A_2,\bm X_2,Y_2),\dotsc$ $(A_n,\bm X_n,Y_n)$ are i.i.d.\ from a
joint distribution $F_0$, where for each subject $i \in [n]:=\{1, 2,
\ldots, n\}$, $A_i$ is the indicator of non-missing response, $\bm
X_i \in \mathscr{X} \subset \R^d$ is a vector of covariates, and
$Y_i \in \mathbb{R}$ is the response (observed only if $A_i =
1$). In other words, we only observe $(A_i, \bm{X}_i, A_iY_i)$ for
$i \in [n]$. Based on the observed data, our goal is to estimate the {\it
  mean response} $\mu := \mathbb{E}_0[Y]$ in the complete data,
where $(A, \bm X, Y) \sim F_0$ and $\E_0$ indicates that the
expectation is taken over the true data generating distribution
$F_0$.

Since some responses or potential outcomes are not observed, the estimand $\mu$ defined above is not identifiable without further assumptions. For the missing data problem, \citet{rubin1976inference} used the term ``missing at random'' (MAR) to describe data that are
missing for reasons related to completely observed variables in the
data set:

\begin{assumption}[Missing at random (MAR)] \label{assump:mar}
  $A \independent Y | \bm X$ under $F_0$.
\end{assumption}

With the additional assumption that no subject is missing or receives treatment/control with probability $1$, the parameter $\mu$ is identifiable from the data.

\begin{assumption}[Overlap] \label{assump:overlap} For $\bm x \in \mathscr{X}$, $e_0(\bm x) := \P_0(A=1|\bm X = \bm x) \in (0, 1]$.
\end{assumption}

Since the seminal works of \citet{rubin1976inference} and
\citet{rosenbaum1983central}, many statistical methods have been developed for the missing data and observational studies problem based on first
estimating the conditional probability $e_0(\bm x)$ (often called the {\it propensity score} in observational studies). One distinguished example is the {\it inverse probability weighting} (IPW) estimator which
dates back to \citet{horvitz1952generalization},
\[
  \hat{\mu}_{\mathrm{IPW}} = \frac{1}{n} \sum_{s=1}^n \frac{A_iY_i}{\hat{e}(\bm X_i)},
\]
where $\hat{e}(\bm X)$ is a sample-estimate of $e_0(\bm X)$. Observe that under
\Cref{assump:mar,assump:overlap},
\begin{equation} \label{eq:ipw-pop}
  \mu = \E_0\left[\frac{AY}{\P_0(A=1|\bm X,Y)}\right] =
  \E_0\left[\frac{AY}{\P_0(A=1|\bm X)}\right] = \E_0\left[\frac{AY}{e_0(\bm X)}\right],
\end{equation}
which implies that $\hat{\mu}_{\mathrm{IPW}}$ can consistently estimate
the mean response $\mu$ if $\hat{e}$ converges to $e_0$. Notice that the first equality in \eqref{eq:ipw-pop} is due to the tower property of
conditional expectation and is always true if the denominator $\P_0(A=1|\bm X,Y)$ is
positive with probability $1$. The second equality in
\eqref{eq:ipw-pop} uses $\P_0(A=1|\bm X,Y) =
\P_0(A=1|\bm X) \ne 0$ by
\Cref{assump:mar,assump:overlap}.

\section{Sensitivity models}
\label{sec:sensitivity-models}


A critical issue of the above approach is that the MAR
assumption is not verifiable using the data because some
responses/potential outcomes are not observed. Therefore, if the
MAR assumption is violated, the IPW estimator $\hat{\mu}_{\mathrm{IPW}}$ is biased and the confidence interval based on it (and any other existing estimator that assumes MAR) does not cover $\mu$ at the nominal rate. In
practice, statisticians often hope the data analyst can use her subject
knowledge to justify the substantive reasonableness of MAR
assumptions \citep{little2014statistical}.
However, speaking realistically, the MAR assumption is perhaps never
strictly satisfied, so the statistical inference is always subject to
criticism of insufficient control of confounding.

To rebut such criticism and make the statistical analysis more credible, a natural question is how sensitive the results are to the violation of the MAR assumption. This is clearly an important question, and, in fact,
sensitivity analysis is often suggested or required for
empirical publications.\footnote{For example, sensitivity analysis is required in
  the Patient-Centered Outcome Research Institute (PCORI) methodology
  standards for handling missing data, see standard MD-4 in
  \url{https://www.pcori.org/research-results/about-our-research/research-methodology/pcori-methodology-standards}. See also \citet{little2012prevention}.}
Many sensitivity analysis methods have been subsequently developed in the missing data and
observational studies problems. In this paper we will consider a sensitivity
model that is closely related to Rosenbaum's sensitivity
model \citep{rosenbaum1987sensitivity,rosenbaum2002observational}. We
refer the reader to
\citet{robins1999association,scharfstein1999adjusting,imbens2003sensitivity,altonji2005selection,hudgens2006causal,vansteelandt2006ignorance,mccandless2007bayesian,vanderweele2011bias,richardson2014nonparametric,ding2016sensitivity}
for alternative sensitivity analysis methods and \Cref{sec:discussion}
for a more detailed discussion.

We begin with a description of the marginal sensitivity model used in this paper, which is
also considered by \citet{tan2006distributional}. To this end, with a slight
abuse of notation, let $e_0(\bm x, y) = \P_0(A=1|\bm X=\bm
x,Y=y)$. Notice that, when showing the IPW estimator is consistent in
\eqref{eq:ipw-pop}, the MAR assumption
\begin{align}\label{eq:e0}
  e_0(\bm x,y) = e_0(\bm x),
\end{align}
is used in the second equality.

When the MAR assumption is violated, equation \eqref{eq:e0} is no
longer valid. It is well well-known that $e_0(\bm x,y)$ is generally not identifiable from the data without the MAR assumption (proof included in the supplement for completeness).
For this reason we shall refer to a user-specified function
$e_0(\bm{x},y)$ as a \emph{sensitivity model}.


\blue{In this paper we consider the following collection of sensitivity models in which the degree of violation of the MAR assumption is quantified by the odds ratio of $e_0(\bm x, y)$ and $e_0(\bm x)$.}

\begin{defn}[Marginal Sensitivity Model
  ] \label{def:our-sens-model}
  Fix a parameter $\Lambda \geq 1$. For the missing data problem, we assume  $e(\bm x, y ) \in \mathcal{E}(\Lambda)$, where
  \begin{equation} \label{eq:our-sens-model}
    \mathcal{E}(\Lambda) = \Big\{ 0 \le e(\bm x, y) \le 1:\,
    \frac{1}{\Lambda} \le \mathrm{OR}(e(\bm x,y), e_0(\bm x)) \leq
    \Lambda, \text{ for all }\bm x \in\mathscr X,y \in \mathbb{R} \Big\},
  \end{equation}
  and $\mathrm{OR}(p_1,p_2) = [p_1/(1-p_1)]/[p_2/(1-p_2)]$ is the
  odds ratio.
\end{defn}

The relationship between this sensitivity model and Rosenbaum's
sensitivity model is examined in \Cref{sec:comp-with-rosenb}.

\begin{remark} \blue{The set $\mathcal{E}(\Lambda)$ becomes larger as $\Lambda$ increases. When $\Lambda = 1$, $\mathcal{E}(1)$ contains the
    singleton $\{e_0(\bm x)\}$ and corresponds to MAR. When $\Lambda =
    \infty$, $\mathcal{E}(\infty)$ contains all functions $e(\bm x, y)$
    that are bounded between $0$ and $1$.}
  To understand \Cref{def:our-sens-model}, it
  can be conceptually easier to imagine that there is an unobserved variable $U$ that ``summarizes'' all unmeasured
  confounding
  , with the relation between $U$ and the
  potential outcomes being unconstrained in any way. This leads to
  an alternative ``added variable'' representation of sensitivity model
  \citep{rosenbaum2002observational}. \citet{robins2002covariance}
  pointed out that it suffices to consider the conditional
  probabilities when $U$ is either one of the potential outcomes. In
  the remainder of the paper we will follow Robins' suggestion to
  simplify the notation.
\end{remark}


\begin{remark}
  It is often convenient to use the logistic representation of the
  marginal sensitivity model \eqref{eq:our-sens-model}. Denote
  \begin{align*}
    g_0(\bm x) = \mathrm{logit}(e_0(\bm x)) = \log \frac{e_0(\bm x)}{1 -
    e_0(\bm x)} \quad \text{and} \quad g_0(\bm x, y) = \mathrm{logit}(e_0(\bm x, y)),
  \end{align*}
  and $h_0(\bm x,y) = g_0(\bm x) - g_0(\bm x, y)$ be the logit-scale
  difference of the observed data selection probability and the complete
  data selection probability.
  If we further introduce the notation $e^{(h)}(\bm x, y)= [1 + \exp(h(\bm x, y) - g_0(\bm x))]^{-1}$, then
  \begin{align}\label{eq:sensemnp}
    \cE(\Lambda)=\{e^{(h)}(\bm x, y): h \in \mathcal H(\lambda)\},~\mathrm{where}~\mathcal{H}(\lambda)=\left\{h: \mathscr X \times \R
    \rightarrow \R ~\text{and}~ ||h||_\infty \le \lambda \right\},
  \end{align}
  and  $\lambda = \log \Lambda$. This shows that the marginal
  sensitivity model puts bound on the $L_{\infty}$-norm of
  $h$. \blue{It is easy to verify that $e^{(h)}(\bm x, y) \to 0$ or
    1 if $h(\bm x, y) \to \pm \infty$.}
\end{remark}

\section{Confidence Interval for the Mean Response}
\label{sec:new-framework}
\label{sec:conf-interv-under}

In this section we construct confidence intervals for the mean
response $\mu$ under the marginal sensitivity models.

\subsection{Confidence interval in sensitivity analysis}
\label{sec:conf-interv-sens}

We begin by a small extension to the definition of marginal sensitivity models. In
\Cref{def:our-sens-model}, a postulated
sensitivity model $e(\bm x,y)$ is compared with the
data-identifiable $e_0(\bm x) =\P_0(A=1|\bm X=\bm x)$. This model is
most appropriate when $e_0(\bm X)$ is estimated non-parametrically
using the data, but such non-parametric estimation is only possible
for low-dimensional problems due to the curse of dimensionality. More
often, for example in propensity score matching or IPW, $e_0(\bm x)$
is estimated by a parametric model. \blue{In this case it is more appropriate
  to compare the postulated sensitivity model $e(\bm x, y)$ with the
  best parametric approximation to $e_0(\bm x)$:
  \begin{align}
    e_{\bm{\beta_0}}(\bm x) & = \argmin_{\bm \beta \in \Theta} \mathrm{KL}\big(\cL_0(A|\bm X=\bm x)\,\big|\big|\, \cL_{\bm \beta}(A|\bm X=\bm x)\big) \nonumber \\
                            &=\argmax_{\bm \beta \in \Theta} \E_0\left[A \cdot \log e_{\bm
                              \beta}(\bm X) + (1-A) \cdot \log (1 - e_{\bm \beta}(\bm X))|\bm
                              X=\bm x \right], \label{eq:best-parametric-e}
  \end{align}
  where $\mathrm{KL(\mu||\nu)}=\sum_{a}\mu(a)\log(\mu(a)/\nu(a))$ stands for the Kullback-Leibler divergence between two discrete distributions $\mu$ and $\nu$, $\cL_0(A|\bm X=\bm x)$ and $\cL_{\bm \beta_0}(A|\bm X=\bm x)$ denote the laws of the random variable $A|\bm X=\bm x$ under $\P_0$ and $\P_{\bm \beta_0}$, respectively, and
  $\{e_{\bm \beta}(\bm x):=\P_{\bm \beta}(A=1|\bm X=\bm x),~\bm \beta
  \in \Theta \subset \R^d\}$ is a family of parametric models for the selection
  probability.} We recommend to use the
logistic regression model
\begin{align}\label{eq:ebetalogistic}
  e_{\bm \beta}(\bm x) = \frac{e^{\bm \beta' \bm x}}{1 +
  e^{\bm \beta'
  \bm x}},~\mathrm{or~equivalently}~g_{\bm \beta}(\bm x)
  = \mathrm{logit}(e_{\bm \beta}(\bm x)) = \bm \beta' \bm x,
\end{align}
which works seamlessly with our sensitivity model because the
degree of violation of MAR is quantified by odds ratio. However our
framework can be easily applied to other parametric models $e_{\bm
  \beta}(\bm x)$ (e.g.\ different links).

\begin{defn}[Parametric Marginal Sensitivity
  Model]\label{defn:psensem} Fix a parameter $\Lambda \geq 1$, this
  collection of sensitivity models assumes the true selection probability $e_0(\bm x,y) = \P_0(A=1|\bm X = \bm x, Y = y)$ satisfies
  \begin{equation}\label{eq:parasensem}
    e_0(\bm x, y) \in \mathcal{E}_{\bm \beta_0}(\Lambda) := \Big
    \{e(\bm x, y):\,\frac{1}{\Lambda} \le \mathrm{OR}(e(\bm x, y), e_{\bm
      \beta_0}(\bm x)) \le \Lambda ~\text{for all } \bm x
    \in\mathscr X, y \in \mathbb{R} \Big\}.
  \end{equation}
  As in \eqref{eq:sensemnp}, the constraint \eqref{eq:parasensem} is equivalent
  to $h_{\bm \beta_0}(\bm x,y) \in \mathcal{H}(\lambda)$ for $h_{\bm \beta}(\bm x,y) = g_{\bm \beta}(\bm x) - g_0(\bm x, y)$ and $\lambda = \log \Lambda$.
\end{defn}

Next we define the shifted estimand under a specific sensitivity
model $h$:
\begin{equation}\label{eq:muh}
  \begin{split}
    \mu^{(h)} & =\left(\E_0 \left[\frac{A}{e^{(h)}(\bm X, Y)}\right] \right)^{-1}\E_0 \left[ AY \left(1 + e^{h(\bm X, Y) - g_{\bm \beta_0}(\bm X)}  \right)\right]  \\
    &=\left(\E_0 \left[\frac{A}{e^{(h)}(\bm X, Y)}\right] \right)^{-1}\E_0 \left[ \frac{AY}{e^{(h)}(\bm X, Y)}\right] \\
  \end{split}
\end{equation}
where $e^{(h)}(\bm x, y)=(1 + e^{h(\bm x, y) - g_{\bm \beta_0}(\bm
  x)})^{-1}$. By the definition of $h_{\bm \beta}$ and the first
equality in \eqref{eq:ipw-pop}, it is easy to see that $\E_0 \left[\frac{A}{e^{(h_{\bm \beta_0})}(\bm X, Y)}\right]=1$ and $\mu^{(h_{\bm \beta_0})}$ are equal to the true mean
response $\mu$. The
set $\{\mu^{(h)} :~e^{(h)}\in\mathcal{E}_{\bm \beta_0}(\Lambda)\}$
will be referred to as
the \emph{partially identifiable region} under $\mathcal{E}_{\bm
  \beta_0}(\Lambda)$; see \Cref{sec:part-ident-param} for more
discussion. \blue{This set tends to the range of $Y$ when $\Lambda \to
  \infty$.}

\begin{remark} \label{rmk:abuse-notation}
  The definitions of $\mu^{(h)}$ and $e^{(h)}$ above can be easily
  extended to the non-parametric marginal sensitivity model by
  replacing $g_{\bm \beta_0}(\bm x)$ by $g_0(\bm x)$. Our statistical
  method can be applied regardless of  the ``baseline'' choice of
  parametric or nonparametric model for $e_0(\bm X)$, as long as the
  model is ``smooth'' enough so that the bootstrap is valid. For simplicity,
  hereafter we will often refer to $h(\bm x, y) \in
  \mathcal{H}(\lambda)$ instead of $e^{(h)}(\bm x, y)$ as
  the sensitivity model, since the former does not depends on the choice of
  the working model for $e_0(\bm X)$. Consequently we will also call
  $\mathcal{H}(\lambda)$ a collection of sensitivity models without
  specifying which parametric/nonparametric model was used for $e_0(\bm X)$.
\end{remark}

\begin{remark} \blue{Compared to \Cref{defn:psensem}, the only
    difference in \eqref{eq:parasensem} is that the postulated
    sensitivity model  $e(\bm x,y)$ is compared to the parametric
    model $e_{\bm \beta_0}(\bm x)$ instead of the nonparametric
    probability $e_0(\bm x)$.} In other words, the parametric sensitivity
  model considers both
  \begin{enumerate}
  \item Model misspecification, that is, $e_{\bm \beta_0}(\bm x) \ne
    e_0(\bm x)$; and
  \item Missing not at random, that is, $e_0(\bm x) \ne e_0(\bm x, y)$.
  \end{enumerate}
  \blue{This is a desirable feature in the sense that the term ``sensitivity
    analysis'' is also widely used as the analysis of an empirical study's
    robustness to parametric modeling assumptions. However, it might also
    make the choice of the sensitivity parameter $\lambda$ more
    difficult in practice. This issue is briefly explored in the
    simulation in \ref{sec:numer-illustr}.}
\end{remark}

With these notations, we can now define what is meant by a confidence interval  under a collection of sensitivity models:

\begin{defn}\label{defn:conf-int-sens-model} A data-dependent
  interval $[L,U]$ is called a confidence interval for
  the mean response $\mu$ with at least $(1-\alpha)$ coverage under the collection of sensitivity models
  $\mathcal{H}(\lambda)$ (may corresponds to $\mathcal{E}(\Lambda)$ or
  $\mathcal{E}_{\bm \beta_0}(\Lambda)$, see \Cref{rmk:abuse-notation}),
  if
  \begin{equation*}
    \label{eq:coverage-definition}
    \P_0(\mu \in  [L,U]) \ge 1 - \alpha
  \end{equation*}
  is true for any data generating distribution $F_0$ such that $h_0
  \in \mathcal{H}(\lambda)$ or $h_{\bm \beta_0} \in
  \mathcal{H}(\lambda)$, depending on whether $\mathcal{E}(\Lambda)$ or
  $\mathcal{E}_{\bm \beta_0}(\Lambda)$ is used. The
  interval is said to be an asymptotic
  confidence interval with at least $(1-\alpha)$ coverage (or strong
  nominal coverage), if $\lim\inf_{n \rightarrow
    \infty}\P_0(\mu \in  [L,U]) \ge 1 - \alpha$.
\end{defn}

\subsection{The IPW Point Estimates}
\label{sec:ipw-point-estimate}

Intuitively, a confidence interval $[L,U]$ as in Definition
\ref{defn:conf-int-sens-model} must at least include a point estimate
of $\mu^{(h)}$ for every $h \in \mathcal{H}(\lambda)$. To this end,
let $h$ be a postulated sensitivity model. The corresponding
selection probability $e^{(h)}(\bm x, y)$ can then be estimated by
\begin{equation}\label{eq:ehat}
  \hat{e}^{(h)}(\bm x,y)= \frac{1}{1 + e^{h(\bm x, y) - \hat g(\bm x)}},
\end{equation}
where
\begin{equation}\label{eq:ghat}
  \hat g (\bm x)=
  \left\{
    \begin{array}{ccc}
      \mathrm{logit}(\hat \P(A=1|\bm X=x))  &  \text{if}  & e^{(h)}(\bm x,y) \in \cE(\Lambda) \\
      \mathrm{logit}(\P_{\hat{\bm \beta}}(A=1|\bm X=x)) = \hat{\bm \beta}' \bm x &      \text{if}  & e^{(h)}(\bm x,y) \in \cE_{\bm \beta_0}(\Lambda).
    \end{array}
  \right.
\end{equation}

The population quantities $\E_0 \left[ \frac{AY}{e^{(h)}(\bm X,
    Y)}\right]$, which is equal to $\mathbb{E}_0[Y]$ for $h = h_0$ or
$h_{\bm \beta_0}$, can be estimated using the IPW estimator:
\[
  \hat{\mu}_{\mathrm{IPW}}^{(h)} = \frac{1}{n} \sum_{i=1}^n
  \frac{A_i Y_i}{\hat{e}^{(h)}(\bm X_i,Y_i)}, \]
where $\hat e^{(h)}$ is as in \eqref{eq:ehat}. However, it is well
known that, even under the MAR assumption, the IPW
estimator can be unstable when the selection probability $e_0(\bm
x)$ (or the parametric approximation $e_{\bm \beta_0}(\bm x)$) is
close to $0$ for some $\bm x\in \sX$
\citep{kang2007demystifying}. To alleviate this issue, the {\it
  stabilized IPW} (SIPW) estimator, obtained by normalizing the
weights, is often used in practice:
\begin{align} \label{eq:sipw}
  \hat{\mu}^{(h)} &= \bigg[\frac{1}{n} \sum_{i=1}^n
                    \frac{A_i}{\hat{e}^{(h)}(\bm X_i,Y_i)}\bigg]^{-1} \bigg[\frac{1}{n} \sum_{i=1}^n
                    \frac{A_i Y_i}{\hat{e}^{(h)}(\bm X_i,Y_i)} \bigg].
\end{align}
It is easy to see that $\hat{\mu}^{(h)}$ estimates $\mu^{(h)}$ defined
in \eqref{eq:muh}.

Compared to IPW, the SIPW estimator is sample bounded \citep{robins2007comment}, that is,
\[ \hat{\mu}^{(h)} \in \left[\min_{i: A_i = 1} Y_i,  \max_{i: A_i=1} Y_i \right]. \]
This property is even more desirable in sensitivity analysis because
$e^{(h)}$ is almost always not the true selection probability, so the
total unnormalized weights $\frac{1}{n}\sum_{i=1}^n A_i/\hat{e}^{(h)}(\bm
X_i,Y_i)$ can be very different from $1$. For this reason, we will use
the SIPW estimator for the remainder of this paper.

Heuristically, the confidence interval $[L,U]$ should at least contain the
range of SIPW point estimates, $\big[\inf_{h \in \mathcal{H}(\lambda)}
\hat{\mu}^{(h)},\sup_{h \in \mathcal{H}(\lambda)}
\hat{\mu}^{(h)}\big]$. We defer the numerical computation of the extrema
of SIPW point estimates till \Cref{sec:line-fract-progr}. For now we
will focus on constructing the confidence interval $[L,U]$ assuming
the range of point estimates can be efficiently computed.

\subsection{Constructing the Confidence Interval}
\label{sec:constr-conf-interv}

To construct a confidence interval for $\mu$, we need to consider
the sampling variability of the SIPW estimator described above. In the
MAR setting, the most common way to estimate the variance is
the asymptotic sandwich formula or the bootstrap
\citep{efron1994introduction,austin2016variance}. In sensitivity
analysis, we also need to
consider all possible violations of the MAR assumption in
$\cH(\lambda)$. In this case, optimizing the estimated
asymptotic variance over $\cH(\lambda)$ is generally computationally
intractable, as explained below.

\subsubsection{The Union Method}
\label{sec:union-method}

We begin by showing how individual confidence intervals of $\mu^{(h)}$
for $h \in H(\lambda)$ can be combined into a confidence interval in
sensitivity analysis.

\begin{ppn} \label{ppn:union-ci}  Suppose there exists data-dependent
  intervals $[L^{(h)}, U^{(h)}]$ such that
  \[
    \lim\inf_{n \rightarrow \infty}\mathbb{P}_0(\mu^{(h)} \in
    [L^{(h)},U^{(h)}]) \ge 1 - \alpha
  \] holds for every $h \in \cH(\lambda)$.
  \begin{itemize}
  \item[(1)] Let $L= \inf_{h \in \mathcal{H}(\lambda)} L^{(h)}$, $U=\sup_{h \in
      \mathcal{H}(\lambda)} U^{(h)}$.
    Then $[L,U]$ is an asymptotic confidence interval of $\mu$ with at
    least $(1-\alpha)$ coverage under the collection of
    sensitivity models $\cH(\lambda)$.
  \item[(2)]  Moreover, if there exists $\alpha' \in [0, \alpha]$ (not depending on $h$) such that
    \begin{equation} \label{eq:extension-alpha-prime}
      \limsup_{n \to \infty} \P_0\big(\mu^{(h)} < L^{(h)}\big) \le
      \alpha' \quad \text{and} \quad \limsup_{n \to \infty} \P_0\big(\mu^{(h)} > U^{(h)}\big) \le
      \alpha - \alpha',
    \end{equation}
    for all $h \in \mathcal{H}(\lambda)$, then the union interval $[L,U]$ covers the partially identified
    region with probability at least $1 - \alpha$,
    \[
      \liminf_{n \to \infty} \P_0\left(\{\mu^{(h)}: h \in \mathcal H(\lambda)\}
        \subseteq [L,U]\right) \ge 1 - \alpha.
    \]
  \end{itemize}
\end{ppn}

\Cref{ppn:union-ci} suggests the following way to construct a
confidence interval for $\mu$ in sensitivity analysis using the
asymptotic distribution of $\hat \mu^{(h)}_1$. Using the general
theory of $Z$-estimation, it is not difficult to establish that $$\sqrt n \left(\hat \mu^{(h)}-\mu^{(h)}\right) \dto N(0, (\sigma^{(h)})^2).$$
See, for example, \citet{lunceford2004stratification} or
\Cref{cor:normality} in the supplement. Then using the sandwich
variance estimator $(\hat{\sigma}^{(h)})^2$, an asymptotically
confidence interval of $\mu^{(h)}$ at least $(1-\alpha)$ coverage is
\[
  [L_{\mathrm{sand}}^{(h)},U_{\mathrm{sand}}^{(h)}] = \left[\hat{\mu}^{(h)}- z_{\frac{\alpha}{2}} \cdot \frac{\hat{\sigma}^{(h)}}{\sqrt n}, \hat{\mu}^{(h)}
    + z_{\frac{\alpha}{2}} \cdot \frac{\hat{\sigma}^{(h)}}{\sqrt n}\right],
\]
where $z_{\frac{\alpha}{2}}$ is the upper
$\frac{\alpha}{2}$-quantile. Then, by Proposition \ref{ppn:union-ci},
an asymptotically confidence interval with at least $(1-\alpha)$ under the
collection of sensitivity models is $[L_{\mathrm{sand}},U_{\mathrm{sand}}]$,
\begin{equation*}
  \begin{split}
    L_{\mathrm{sand}} = \inf_{h \in \mathcal{H}(\lambda)} L_{\mathrm{sand}}^{(h)}, \quad
    U_{\mathrm{sand}} = \sup_{h \in \mathcal{H}(\lambda)} U_{\mathrm{sand}}^{(h)}.
  \end{split}
\end{equation*}
However, the standard error $\hat{\sigma}^{(h)}$ is a very complicated function of
$h$ (see \Cref{cor:normality}) and numerical optimization over $h \in
\cH(\lambda)$ is practically infeasible.

\subsubsection{The Percentile Bootstrap}
\label{sec:nonp-bootstr}

The centerpiece of our proposal is to use the percentile bootstrap to
construct $[L^{(h)},U^{(h)}]$. Next we introduce the necessary
notations to describe the bootstrap procedure. Let $\P_n$ be the
empirical measure of the sample $\bm T_1, \bm T_2, \ldots, \bm T_n$,
where $\bm T_i=(A_i, \bm X_i', A_iY_i)$, and $\hat {\bm T}_1, \hat
{\bm T}_2, \ldots, \hat {\bm T}_n$ be i.i.d. resamples from the
empirical measure. Let $\hat{\hat{\mu}}^{(h)}$ the SIPW estimate
\eqref{eq:sipw} computing using the bootstrap resamples $\{\hat{\bm
  T_i}\}_{i\in[n]}$. For $h \in
\cH(\lambda)$, the percentile bootstrap confidence interval of
$\mu^{(h)}$ is given by
\begin{align}
  \label{eq:lbh}
  [L^{(h)},U^{(h)}] =\left[Q_{\frac{\alpha}{2}}(\hat{\hat{\mu}}^{(h)}), Q_{1-\frac{\alpha}{2}}(\hat{\hat{\mu}}^{(h)})\right],
\end{align}
where $Q_{\alpha}(\hat{\hat{\mu}})$ is the $\alpha$-percentile of $\hat{\hat{\mu}}$ in the bootstrap distribution, that is,
$$Q_{\alpha}(\hat{\hat{\mu}}):=\inf\{t: \hat \P_n(\hat{\hat{\mu}} \leq t) \geq \alpha\},$$
where $\hat \P_n$ is the bootstrap resampling distribution.

%
%

We begin by showing that the percentile bootstrap interval $[L^{(h)}, U^{(h)}]$ is an asymptotically valid confidence interval of $\mu^{(h)}$
for the parametric sensitivity model $e^{(h)} \in \cE_{\beta_0}(\Lambda)$.

\begin{thm}[Validity of the Percentile
  Bootstrap] \label{thm:valid-pbz} In the logistic model
  \eqref{eq:ebetalogistic} and under regularity assumptions
  (see Assumption \ref{assum:moment} in the supplement), for every $e^{(h)} \in \cE_{\bm \beta_0}(\Lambda)$
  we have
  $$\limsup_{n \to \infty}\mathbb{P}_0\left(\mu^{(h)} < L^{(h)}\right) \le
  \frac{\alpha}{2} \quad \text{and} \quad \limsup_{n \to \infty}\mathbb{P}_0\left(\mu^{(h)} > U^{(h)}\right) \le
  \frac{\alpha}{2}.$$
\end{thm}

The proof of this result, which invokes the general theory of
bootstrap for $Z$-estimators \citep[Chapter 10]{bootz,korosok}, is
explained in Appendix \ref{sec:proofs} in the supplement.

Our percentile bootstrap confidence interval under the collection of
sensitivity models $\cE_{\bm \beta_0}(\Lambda)$ is given by $[L, U]$
where
\begin{equation} \label{eq:pb-cin}
  L = Q_{\frac{\alpha}{2}}\Big(\inf_{h \in \mathcal{H}(\lambda)}
  \hat{\hat{\mu}}^{(h)}\Big) \quad \text{and} \quad U = Q_{1-\frac{\alpha}{2}}\Big(\sup_{h \in \mathcal{H}(\lambda)} \hat{\hat{\mu}}^{(h)}_{b} \Big).
\end{equation}
The important thing to observe here is that the infimum/supremum is
inside the quantile function in \eqref{eq:pb-cin}, which makes
the computation especially efficient using linear programming
(see Section \ref{sec:line-fract-progr}). The interchange of quantile
and infimum/supremum is justified in the following generalized (von
Neumann's) minimax/maximin inequalities (see
\citet{cohen2012generalized} for a similar result for finite sets).

\begin{lem} \label{lem:minimax}
  Let $L, U$ be as defined in \eqref{eq:pb-cin} and $L^{(h)}, U^{(h)}$
  be as defined in \eqref{eq:lbh}. Then
  \[L \le \inf_{h \in \mathcal{H}(\lambda)} L^{(h)} \quad \textrm{and} \quad   U \ge \sup_{h \in \mathcal{H}(\lambda)} U^{(h)}.\]
\end{lem}

Asymptotic validity of the confidence interval $[L, U]$ in
\Cref{eq:pb-cin} then
immediately follows from the validity of the union method
(\Cref{ppn:union-ci}), the validity of the percentile bootstrap
(\Cref{thm:valid-pbz}), and \Cref{lem:minimax}.

\begin{thm} \label{thm:valid-pb}
  Under the same assumptions as in Theorem \ref{thm:valid-pbz},
  $[L,U]$ is an asymptotic confidence interval of the
  mean response $\mu$ with at least $(1-\alpha)$ coverage, under the collection of sensitivity models
  $\cE_{\bm \beta_0}(\Lambda)$. Furthermore,
  $[L,U]$ covers the partially identified region
  $\{\mu^{(h)}: e^{(h)} \in \mathcal E_{\bm \beta_0}(\Lambda)\}$ with
  probability at least $1 - \alpha$.
\end{thm}

\subsection{Range of SIPW Point Estimates: Linear Fractional Programming}
\label{sec:line-fract-progr}

Theorem \ref{thm:valid-pb} transformed the sensitivity analysis
problem to computing the extrema of the SIPW point estimates, $\inf_{h \in
  \mathcal{H}(\lambda)} \hat{\hat{\mu}}^{(h)}_{b}$ and $\sup_{h \in
  \mathcal{H}(\lambda)} \hat{\hat{\mu}}^{(h)}_{b}$. In practice, we
only repeat this over $B~(\ll n)$ random resamples and compute the
interval by
\begin{equation} \label{eq:pb-ci}
  L_B = Q_{\frac{\alpha}{2}}\bigg(\Big(\inf_{h \in \mathcal{H}(\lambda)} \hat{\hat{\mu}}^{(h)}_b\Big)_{b \in [B]}\bigg), \quad U_B = Q_{1-\frac{\alpha}{2}}\bigg(\Big(\sup_{h \in \mathcal{H}(\lambda)} \hat{\hat{\mu}}^{(h)}_b\Big)_{b \in [B]}\bigg).
\end{equation}
For notational simplicity, below we consider how to compute the extrema $\inf_{h \in
  \mathcal{H}(\lambda)} \hat{{\mu}}^{(h)}$ and $\sup_{h \in
  \mathcal{H}(\lambda)} \hat{{\mu}}^{(h)}$ using the full observed data
instead of the resampled data. \blue{This also gives an interval of point
estimates under the sensitivity model that is shorter than the
confidence interval. We recommend reporting both intervals in any real
data analysis, see \Cref{sec:numer-illustr} for some examples.}

Recalling \eqref{eq:ghat} and
\eqref{eq:sipw}, computing the extrema is equivalent to solving
\begin{equation} \label{eq:lfpI}
  \mathrm{min.~or~max.~}\frac{\sum_{i=1}^n A_i Y_i \left(1 +
      z_i e^{-\hat{g}(\bm X_i)}\right)}{\sum_{i=1}^n A_i \left(1 + z_i
      e^{-\hat{g}(\bm X_i)}\right)}  \text{ subject to } z_i \in
  \left[\Lambda^{-1}, \Lambda\right],\text{ for all } i \in [n],
\end{equation}
where the optimization variables are $z_i=e^{h(\bm X_i,Y_i)}$ for $i
\in [n]$. All the other variables are observed or can be estimated
from the data. Notice that $\hat{g}(\bm x)$ needs to be re-estimated
in every bootstrap resample.
Without loss of generality, assume that the first $1 \le m < n$ responses are observed, that is $A_1 = A_2 = \cdots A_m = 1$ and $A_{m+1} = \cdots =A_n = 0$, and suppose that the observed responses are in decreasing order, $Y_1
\ge Y_2 \ge \cdots \ge Y_m$. Then  \eqref{eq:lfpI} simplifies to
\begin{equation} \label{eq:lfpII}
  \mathrm{min.~or~max.~} \frac{\sum_{i=1}^m Y_i \left(1 + z_i e^{-\hat{g}(\bm X_i)}\right)}{\sum_{i=1}^m \left(1 + z_i e^{-\hat{g}(\bm X_i)}\right)}, \text{ subject to } z_i \in \left[\Lambda^{-1}, \Lambda\right], ~1\leq i \leq m.
\end{equation}
This optimization problem is the ratio of two linear functions of the
decision variables $\bm z$, hence called a {\it linear fractional programming}. It can be transformed to linear programming by the Charnes-Cooper transformation \citep{charnes1962programming}. Denote
\[ \bar{z}_i = \frac{z_i}{\sum_{i=1}^m (1 + z_i e^{-\hat{g}(\bm X_i)})}, \text{ for } 1\leq s \leq m, \text{ and } t= \frac{1}{\sum_{i=1}^m (1 + z_i e^{-\hat{g}(\bm X_i)})}.\]
This translates \eqref{eq:lfpII} into the following linear programming:
\begin{align} \label{eq:lp}
  \mathrm{minimize~or~maximize}\quad&\sum_{i=1}^m Y_i e^{-\hat{g}(\bm X_i)} \bar{z}_i + t \bigg(\sum_{i=1}^m Y_i \bigg)  \nonumber \\
  \mathrm{subject~to}\quad & t \ge 0,~  \Lambda^{-1} t \le \bar{z}_i \le
                             \Lambda t, \text{ for } 1 \leq i \leq m, \nonumber \\
                                    &\sum_{i=1}^m e^{-\hat{g}(\bm X_i)} \bar{z}_i + t \bigg(\sum_{i=1}^m Y_i\bigg) = 1.
\end{align}
Therefore, the range of the SIPW point estimate can be computed
efficiently by solving the above linear program. Furthermore,
the following result shows that the solution of \eqref{eq:lp} must have
the same or opposite order as the outcomes $\bm Y$, which enables even
faster computation of \eqref{eq:lp}.

\begin{ppn} \label{ppn:lfp} Suppose $(z_i)_{i=1}^m$ solves the maximization problem in \eqref{eq:lp}. Then $(z_i)_{i=1}^m$ has the same order as $(Y_i)_{i=1}^m$, that is, if  $Y_{s_1} > Y_{s_2}$ then $z_{s_1} > z_{s_2}$, for $1 \leq s_1\ne s_2 \leq m$. Furthermore, there exists a solution $(z_i)_{i=1}^m$ and a threshold $M$ such that $z_i =\Lambda$, if $Y_i \ge M$, and $z_i =\frac{1}{\Lambda}$, if $Y_i < M$. The same conclusion holds for the minimizer in \eqref{eq:lp} with $(Y_i)_{i=1}^m$ replaced by $(-Y_i)_{s =1}^m$.
\end{ppn}

\Cref{ppn:lfp} implies that we only need to compute the objective of
\eqref{eq:lfpII} for at most $m$ choices of $(z_i)_{i\in[m]}$ by
enumerating the index where it changes from $\Lambda$ to
$\Lambda^{-1}$:
\begin{equation} \label{eq:m-candidates}
  \left\{(z_i)_{i\in[m]}: \exists~  a \in [m] \text{ with } z_i =\Lambda, \text{ for } 1\leq i \leq a, \text{ and } z_i =\frac{1}{\Lambda}, \text{ for } a+1 \leq i \leq m \right\}.
\end{equation}
It is easy to see that we only need $O(m)$ time  to compute
$\sum_{i=1}^m Y_i(1+z_ie^{-\hat{g}(\bm X_i)})$ and $\sum_{i=1}^m
(1+z_ie^{-\hat{g}(\bm X_i)})$, for all the $m$ choices in
\eqref{eq:m-candidates}. Hence, the computational complexity to solve
the linear fractional programming \eqref{eq:lfpII} is $O(m)$. This is
the best possible rate since it takes $O(m)$ time to just compute the
objective once.

The above discussion suggests that the interval $[L_B, U_B]$, which
entails finding $\inf_{h \in \mathcal{H}(\lambda)}
\hat{\hat{\mu}}^{(h)}_{b}$ and $\sup_{h \in \mathcal{H}(\lambda)}
\hat{\hat{\mu}}^{(h)}_{b}$ for every $1 \leq b \leq B$, can be
computed extremely efficiently. The computational complexity is $O(n
B+n \log n)$ ignoring the time spent to fit the logistic propensity
score models, where the
extra $n \log n$ is needed for sorting the data. Indeed, even under
the MAR assumption, we need $O(nB)$ time to compute the Bootstrap
confidence interval of $\mu$. In conclusion, our proposal requires
almost no extra cost to conduct a sensitivity analysis for the IPW
estimator than to obtain its bootstrap confidence interval under MAR.



\section{Confidence Interval for the ATE in the Sensitivity Model}
\label{sec:sens-analys-observ}

\blue{The framework developed above can be easily extended to observational studies, which is essentially two missing data problems. In observational studies, we observe i.i.d.\ $(A_1,\bm X_1,Y_1),\allowbreak (A_2,\bm X_2,Y_2), \ldots, (A_n,\bm X_n,Y_n)$, where for each subject $i \in [n]$, $A_i$ is a binary treatment indicator (which is   $1$ if treated and $0$ in control), $\bm X_i \in \mathscr{X} \subset \R^d$ is a vector of measured confounders, and $$Y_i = Y_i(A_i) =A_iY_i(1) + (1 -A_i)Y_i(0)$$ is the outcome. Furthermore, we assume  $(A_i,\bm{X}_i, Y_i(0),Y_i(1))$ are i.i.d.\ from a joint distribution  $F_0$. Here, we are using \citet{rubin1974estimating}'s potential outcome notation and have assumed the stable unit treatment value  assumption \citep{rubin1980randomization}. The goal is to estimate the {\it average treatment effect} (ATE), $\Delta := \E_0[Y(1)] -  \E_0[Y(0)]$. With the potential outcome notation, the observational studies problem is essentially two missing data problems: use $(A_i,\bm
  X_i,A_iY_i) = (A_i,\bm X_i,A_iY_i(1))$ to estimate $\mu(1) := \E_0[Y(1)]$, and use $(1 - A_i,\bm X_i,(1 - A_i)Y_i) = (1 - A_i,\bm X_i,(1 - A_i)Y_i(0))$ to estimate $\mu(0) := \E_0[Y(0)]$.

  Here, the MAR assumption is replaced by the assumption of strong ignorability or no unmeasured confounder (NUC), that is, $A \independent (Y(0),Y(1))|\bm X$ under $F_0$. The NUC assumption in the observational studies problem implies that $e_{a}(\bm x,y) = e_{a}(\bm x)$, where $e_{a}(\bm x,y) := \P_0(A=1|\bm X=\bm x,Y(a)=y)$, for $a \in \{0, 1\}$. Similarly, the overlap condition (\Cref{assump:overlap}) becomes  $e_0(\bm x) \in (0, 1)$. Under these assumptions, the IPW estimate
  \[
    \quad  \hat{\Delta}_{\mathrm{IPW}} = \frac{1}{n} \sum_{s=1}^n
    \frac{A_iY_i}{\hat{e}(\bm X_i)} - \frac{(1 - A_i)Y_i}{1 - \hat{e}(\bm X_i)},
  \]
  can consistently estimate the ATE $\Delta$, where $\hat{e}(\bm X)$ is a sample-estimate of $e_0(\bm X)$.}

As before, suppose we use a parametric model
$e_{\bm \beta_0}(\bm{x})$ to model the propensity score $\P_0(A=1|\bm
X = \bm x)$ using the observed covariates. The corresponding
parametric marginal sensitivity model assumes
\begin{equation}\label{eq:prIIobs}
  \frac{1}{\Lambda} \le \mathrm{OR}(e_{a}(\bm x,y), e_{\bm \beta_0}(\bm
  x)) \leq \Lambda, \text{ for all }\bm x \in\mathscr X,y \in
  \mathbb{R}, a \in \{0,1\}.
\end{equation}
\blue{One can analogously define the difference between $e_{a}(\bm x,y)$ and
$e_{\bm \beta_0}(\bm x)$ in the logit scale and rewrite
\eqref{eq:prIIobs} as an $L_{\infty}$-constraints on the
difference. Similarly, one can define the shifted propensity score and
the shifted ATE. We omit the details for brevity but would like to
mention that the shifted estimand $\Delta^{(h_0,h_1)}$ now depends on
two $h$ functions corresponding to the two potential outcomes. The
SIPW estimator of $\Delta^{(h_0,h_1)}$ can be defined in the same way
as \eqref{eq:sipw}.}

Now, just as in Section \ref{sec:nonp-bootstr}, we can use the percentile
bootstrap obtain a asymptotically valid interval for
$\Delta^{(h_0,h_1)}$:
\[
  \left[Q_{\frac{\alpha}{2}}(\hat{\hat{\Delta}}^{(h_0,h_1)}), Q_{1-\frac{\alpha}{2}}(\hat{\hat{\Delta}}^{(h_0,h_1)})\right],
\]
where $h_0$ and $h_1$ are held fixed and $Q_{\frac{\alpha}{2}}(\hat{\hat{\Delta}}^{(h_0,h_1)})$ is the
$\alpha$-th bootstrap quantile of the SIPW estimates. Then, by
interchanging the maximum/minimum and the quantile as in Theorem
\ref{cor:valid-pb-obs}, we obtain a confidence interval for the
ATE:

\begin{cor} \label{cor:valid-pb-obs}
  Under the same assumptions in Theorem \ref{thm:valid-pbz}, the
  confidence interval
  \begin{align}\label{eq:sen_interval_obs}
    \left[ Q_{\frac{\alpha}{2}}\left(\inf_{h_0,h_1 \in \mathcal{H}(\lambda)} \hat{\hat{\Delta}}^{(h_0,h_1)}\right), Q_{1-\frac{\alpha}{2}}\left(\inf_{h_0,h_1 \in \mathcal{H}(\lambda)} \hat{\hat{\Delta}}^{(h_0,h_1)} \right) \right]
  \end{align}
  covers the average treatment effect $\Delta$ is with probability at
  least $(1-\alpha)$ asymptotically under the collection of parametric sensitivity models \eqref{eq:prIIobs}.
\end{cor}


The interval in \eqref{eq:sen_interval_obs} can be computed
efficiently using linear fractional programming as in Section
\ref{sec:line-fract-progr}. To simplify notation, assume, without loss
of generality, the first $m \leq  n$ units are treated ($A = 1$) and
the rest are the control ($A=0$), and that the outcomes are ordered
decreasingly among the first $m$ units and the other $n-m$
units. Then, as in \eqref{eq:lfpII}, computing the interval
\eqref{eq:sen_interval_obs} is equivalent to solving the following
optimization problem:
\begin{equation} \label{eq:lfp-ate}
  \begin{split}
    \mathrm{minimize~or~maximize}\quad&\frac{\sum_{i=1}^m Y_i (1 + z_i
      e^{-\hat{g}(\bm X_i)})}{\sum_{i=1}^m (1 + z_i
      e^{-\hat{g}(\bm X_i)})} -  \frac{\sum_{i=m+1}^n Y_i (1 + z_i
      e^{\hat{g}(\bm X_i)})}{\sum_{i=m+1}^n (1 + z_i
      e^{\hat{g}(\bm X_i)})} \\
    \mathrm{subject~to}\quad& \frac{1}{\Lambda} \le z_i \le
    \Lambda, \text{ for } 1 \leq i \leq n,
  \end{split}
\end{equation}
where $z_i=e^{h_1(\bm X_i, Y_i)}$, for $1 \leq i \leq m$ and $z_i=e^{-h_0(\bm X_i, Y_i)}$, for $m+1 \leq i \leq n$. Note that the variables $(z_i)_{i=1}^m$ and $(z_i)_{i=m+1}^n$ are
separable in \eqref{eq:lfp-ate}, so we can solve the maximization/minimization problem in \eqref{eq:lfp-ate} by solving one maximization/minimization problem for $(z_i)_{i=1}^m$ and one minimization/maximization problem for $(z_i)_{i=m+1}^n$.
Therefore, similar to the missing data problem, the time complexity to
obtain the range of the SIPW estimates $\hat{\Delta}$, over a range of $B$ bootstrap resamples, is only $O(nB+n \log n)$.


\section{Extensions}
\label{sec:extension}

In this section we discuss three extensions of the general framework described in \Cref{sec:new-framework}.

\subsection{Mean of the Non-Respondents and Average Treatment Effect on the Treated}
\label{sec:mean-non-respondents}

In many applications, it is also interesting to estimate the {\it mean of the non-respondents} $\mu_{0}=\E_0[Y | A = 0]$, and the {\it average treatment effect on the treated} (ATT) $\Delta_1 =
\E_0[Y(1) - Y(0)|A=1]$. The method described in \Cref{sec:new-framework} can be easily applied to these estimands, as described below.

To begin with note that
\[
  \mu_{0} = \mathbb{E}_0[Y | A = 0] = \frac{\E_0[Y\cdot 1\{A=0\}]}{\P_0(A=0)}= \frac{\E_0\left[ (1 - e_0(\bm X,Y)) Y \right]}{\P_0(A=0)} = \frac{\E_0\left[\frac{1 - e_0(\bm X,Y)}{e_0(\bm X,Y)} A Y \right]}{\P_0(A=0)}.\]
Then as in \Cref{thm:valid-pb}, under the collection of parametric sensitivity models $\mathcal E_{\bm \beta_0}(\Lambda)$,
\begin{align}\label{eq:sen_interval_obs_nr}
  \left[ Q_{\frac{\alpha}{2}}\bigg(\Big(\inf_{h \in \mathcal{H}(\lambda)} \hat{\hat{\mu}}^{(h)}_{0b} \Big)_{b \in [B]}\bigg),~ Q_{1-\frac{\alpha}{2}}\bigg(\Big(\inf_{h \in \mathcal{H}(\lambda)} \hat{\hat{\mu}}^{(h)}_{0b} \Big)_{b \in [B]}\bigg) \right],
\end{align}
is an asymptotic confidence interval of the
non-respondent mean $\mu_0$ with at least $(1-\alpha)$ coverage, where the SIPW estimate is
$$\hat{\mu}_{0}^{(h)} = \frac{\sum_{i=1}^n e^{h(\bm X_i,Y_i) - \hat{g}(\bm X_i)} A_i Y_i}{\sum_{i=1}^n e^{h(\bm X_i,Y_i)
    - \hat{g}(\bm X_i)} A_i},$$
where $\hat g$ is as in \eqref{eq:ghat} and $\hat{\hat{\mu}}^{(h)}_{01}, \hat{\hat{\mu}}^{(h)}_{02}, \ldots, \hat{\hat{\mu}}^{(h)}_{0B}$ are the $B$ bootstrap resamples of $\hat{\mu}_{0}^{(h)}$.
As before, the interval \eqref{eq:sen_interval_obs_nr} can be computed efficiently using linear fractional programming. In particular, it is easy to verify that
\Cref{ppn:lfp} still holds and thus we only need to consider the $O(n)$
candidate solutions in \Cref{eq:m-candidates}.



For the ATT, notice that $\Delta_{1} = \E[Y(1)|A=1] -
\E[Y(0)|A=1]$. The first term is identifiable using the data and the
second term is the non-respondent mean by treating $Y(0)$ as the response. We can use the same procedure in \Cref{sec:sens-analys-observ} to obtain confidence interval of
$\Delta_{1}$ under $\mathcal E_{\bm \beta_0}(\Lambda)$, using the
analogously defined SIPW estimates $\hat \Delta_1^{(h_0)}$.

\subsection{Augmented Inverse Probability Weighting}
\label{sec:augm-inverse-prob}

Besides the basic IPW and SIPW estimators, another commonly used
estimator in missing data and observational studies is the {\it
  augmented inverse probability weighting} (AIPW) estimator which has a double robustness property explained below \citep{robins1994estimation}. As usual, we consider the missing data
problem first for simplicity. Apart from the selection probability, the AIPW estimator also utilizes another nuisance parameter, $f_0(\bm x) = \mathbb{E}_0[Y|A=1,\bm X = \bm x]$. Suppose this is estimated by $\hat{f}(\bm x)$ from the sample, for example, by linear regression, and $e_0(\bm x)$ is estimated by $\hat e(\bm x)$. Then the AIPW estimator (with weight stabilization)
is given by
\[
  \hat \mu_{\mathrm{AIPW}} = \frac{1}{n} \sum_{i=1}^n \frac{A_i
    Y_i}{\hat{e}(\bm X_i)} - \frac{A_i -
    \hat{e}(\bm X_i)}{\hat{e}(\bm X_i)}
  \hat{f}(\bm X_i).
\]
Let $\bar{e}(\bm x)$ and $\bar{f}(\bm x)$ be the large-sample limits of $\hat{e}(\bm x)$ and
$\hat{f}(\bm x)$. When $e_0(\bm x)$ and $f_0(\bm x)$ are estimated non-parametrically, then $\bar{e}(\bm x) = e_0(\bm x)$ and $\bar{f}(\bm x) = f_0(\bm x)$. When $e_0(\bm x)$ and $f_0(\bm x)$ are estimated parametrically, $\bar{e}(\bm x) =e_{\bm \beta_0}(\bm x)$ and $\bar{f}(\bm x) =f_{\bm{\theta}_0}(\bm X)$, the best parametric approximations (see \eqref{eq:best-parametric-e}). Then it is easy to show that
$\hat \mu_{\mathrm{AIPW}}$ always estimates (regardless of the
correctness of MAR)
\begin{equation} \label{eq:aipw-limit}
  \hat \mu_{\mathrm{AIPW}} \overset{P}{\to} \E_0[Y] +
  \E_0\left[\frac{A -
      \bar{e}(\bm X)}{\bar{e}(\bm X)} (Y -
    \bar{f}(\bm X))\right].
\end{equation}
Under MAR (\Cref{assump:mar}), by first taking expectation
conditioning on $\bm X$ it is straightforward to show that the second
term is $0$ if $\bar{e}(\bm x) =e_0(\bm X)$ or $\bar{f}(\bm x) = f_0(\bm x)$. In other
words, $\hat \mu_{\mathrm{AIPW}}$ is consistent for $\mu$ if MAR
holds and at least one of $\hat{e}(\bm x)$ and
$\hat{f}(\bm x)$ is consistent, a property called \emph{double
  robustness}.

When MAR does not hold, by taking expectation conditioning on
$\bm X$ and $Y$, \eqref{eq:aipw-limit} implies that
\[
  \hat \mu_{\mathrm{AIPW}} - \mu \overset{P}{\to} \E_0\left[\frac{e_0(\bm X,Y) - \bar{e}(\bm X)}{\bar{e}(\bm X)} (Y -
    \bar{f}(\bm X)) \right].
\]
Now, as in \Cref{sec:new-framework}, we consider the collection of
parametric sensitivity models $\mathcal{E}_{\bm \beta_0}(\Lambda)$ so
$\bar{e}(\bm x) = e_{\bm \beta_0}(\bm x)$. For
$h \in \mathcal{H}(\lambda)$, define
\[
  \tilde{\mu}^{(h)}_{\bar{f}} = \mu + \E_0\left[
    \frac{e_0(\bm X,Y) -
      e^{(h)}(\bm X,Y)}{e^{(h)}(\bm X,Y)} (Y -
    \bar{f}(\bm X)) \right],
\]
where $e^{(h)}(\bm x, y)=\big[1 + e^{h(\bm x, y) -
  \mathrm{logit}(\P_0(A=1|\bm X=x))} \big]^{-1}$.  As before, it is
obvious that $\tilde{\mu}^{(h_{\bm \beta_0})}_{\bar{f}} = \mu$, the true mean response,
where $h_{\bm \beta_0}(\bm x,y) =\mathrm{logit}(e_{\bm \beta_0}(\bm x)) - \mathrm{logit}(e_{0}(\bm x,y))$.

The AIPW estimator of $\tilde{\mu}^{(h)}_{\bar{f}} $ is
\begin{equation} \label{eq:aipw}
  \begin{split}
    \hat{\mu}^{(h)}_{\mathrm{AIPW}} &= \frac{1}{n} \sum_{i=1}^n \frac{A_i
      Y_i}{\hat{e}^{(h)}(\bm X_i)} - \frac{A_i -
      \hat{e}^{(h)}(\bm X_i)}{\hat{e}^{(h)}(\bm X_i)}
    \hat{f}(\bm X_i) \\
    &= \frac{1}{n} \sum_{i=1}^n \hat{f}(\bm X_i) + \frac{1}{n}
    \sum_{i=1}^n \frac{A_i(Y_i -
      \hat{f}(\bm X_i))}{\hat{e}^{(h)}(\bm X_i)},
  \end{split}
\end{equation}
where $\hat{e}^{(h)}(\bm x,y)= \big[1 + e^{h(\bm x, y) - \hat g(\bm x,
  y)}\big]^{-1}$ and where $\hat g$ is as in  \eqref{eq:ghat}. The second term on the right hand side of \eqref{eq:aipw}
is not sample bounded, so it
is often preferable to use the stabilized weights. This results in
the following stabilized AIPW (SAIPW) estimator:
\begin{equation} \label{eq:saipw}
  \begin{split}
    \hat{\mu}^{(h)}_{\mathrm{SAIPW}} &= \frac{1}{n} \sum_{i=1}^n \hat{f}(\bm X_i) + \frac{1}{\frac{1}{n} \sum_{i=1}^n A_i\hat{e}^{(h)}(\bm X_i) } \left[\frac{1}{n} \sum_{i=1}^n \frac{A_i(Y_i - \hat{f}(\bm X_i))}{\hat{e}^{(h)}(\bm X_i)} \right] \\
    &= \frac{1}{n} \sum_{i=1}^n \hat{f}(\bm X_i) + \frac{\sum_{i=1}^n A_i(Y_i - \hat{f}(\bm X_i)) (1 + e^{h(\bm X_i,Y_i) - \hat{g}(\bm X_i)})}{\sum_{i=1}^n A_i (1 +  e^{h(\bm X_i,Y_i) - \hat{g}(\bm X_i)})}.
  \end{split}
\end{equation}
As before, this estimates $\mu$ when $h = h_{\bm \beta_0}$.

Compared to the SIPW estimator \eqref{eq:sipw}, in \eqref{eq:saipw} we
replace the response $Y_i$ by $Y_i - \hat{f}(\bm X_i)$ and add an
offset term $\frac{1}{n}\sum_{i=1}^n\hat{f}(\bm X_i)$. Therefore,
computing the extrema of \eqref{eq:saipw} can still be formulated as
linear fractional programming and the numerical computation is
still efficient. To construct asymptotically valid confidence intervals, the outcome regression model $\hat{f}(\bm X_i)$ must be
parametric (for example, linear regression). The
$Z$-estimation framework in \Cref{sec:proofs} in the supplement can then be extended to show
that \Cref{thm:valid-pbz} (validity of percentile bootstrap) still
holds for SAIPW.

Similar to \Cref{sec:sens-analys-observ,sec:extension}, the above procedures can be extended to observational studies for estimating the ATE $\Delta$, the non-respondent mean $\mu_0$, or the ATT $\Delta_1$, using analogously defined SAIPW estimates  $\hat \Delta_{\mathrm{SAIPW}}$, $\hat \mu_{0, \mathrm{SAIPW}}$, and $\hat \Delta_{1, \mathrm{SAIPW}}$, respectively.

\subsection{Lipschitz Constraints in the Sensitivity Model}
\label{sec:lipsch-senst-model}

So far we have focused on the marginal sensitivity models,
\begin{align*}
  \cE(\Lambda)~\text{or}~\cE_{\bm \beta_0}(\Lambda)=\{e^{(h)}(\bm x, y): h \in \mathcal H(\lambda)\}.
\end{align*}
Although this model is very easy to interpret, some deviations $h \in \mathcal{H}(\lambda)$
may be deemed unlikely because the function $h$ is not smooth. Here, we consider an extension of our sensitivity model which assumes $h$ is also Lipschitz-continuous.
Formally, define
\begin{align*}
  \cE(\Lambda)~\text{or}~\cE_{\bm \beta_0}(\Lambda):=\{e^{(h)}(\bm x, y): h \in \mathcal H_{\lambda, L}\},
\end{align*}
where  $$\mathcal{H}_{\lambda,L} = \mathcal{H}(\lambda) \bigcap \left\{h: \frac{|h(\bm x_1,y_1) - h(\bm x_2,y_2)|}{d((\bm x_1,y_1), (\bm x_2,y_2))} \le L, \text{ for all } \bm x_1, \bm x_2 \in \sX, \text{ and } y_1, y_2 \in \R \right\},$$
and $d$ is a distance metric defined on the space $\mathscr{X} \times
\mathbb{R}$ and $L > 0$ is the Lipschitz constant.

As $\mathcal H_{\lambda, L} \subseteq \mathcal H(\lambda)$, the validity
of the percentile bootstrap obviously holds for functions in $\mathcal
H_{\lambda, L}$. To obtain a range of point estimates and confidence intervals under $\cE_{\bm \beta_0}(\Lambda, L)$, we just need to add the following $n(n-1)$ constraints in the optimization problem \eqref{eq:lfpI}:
\[
  z_i \le e^{L\cdot d((\bm X_i, Y_i), (\bm X_j, Y_j))} z_j, \text{ for all }   1\leq i \ne j  \leq n,
\]
These constraints are linear in $\{z_i\}_{i=1}^m$ and the resulting optimization problem is still a linear fractional program, which can be efficiently computed. However, \Cref{ppn:lfp} no longer holds, so we cannot use the algorithm described after \Cref{ppn:lfp} to solve the optimization problem corresponding to $\mathcal{H}_{\lambda,L}$.

\section{Discussion}
\label{sec:discussion}

\subsection{Related Frameworks of Sensitivity Analysis}
\label{sec:related-work}

If there is no unmeasured confounder, the
potential outcome $Y(a)$ is independent of the treatment $A$ given
$\bm X$, $Y(a) \independent A | \bm X$, for $a \in \{0, 1\}$. Existing sensitivity analysis
methods have considered at least three types of relaxations of this assumption:
\begin{enumerate}
\item Pattern-mixture models consider a \emph{specific} difference between the conditional
  distribution $Y(a)|\bm X, A$ and $Y(a)|\bm X$
  \citep[e.g.][]{robins1999association,robins2002covariance,birmingham2003pattern,vansteelandt2006ignorance,daniels2008missing}.
\item Selection models consider a \emph{specific} difference between the conditional
  distribution $A|\bm X, Y(a)$ and $A | \bm X$
  \citep[e.g.][]{scharfstein1999adjusting,gilbert2003sensitivity,gilbert2013sensitivity}.
\item Rosenbaum's sensitivity models consider a \emph{range} of
  possible selection models so that within a matched set the
  probabilities of getting treated are no more different than a number
  $\Lambda$ in odds ratio. A worst case p-value of Fisher's sharp null
  hypothesis is then reported \citep[e.g.][Chapter 4]{rosenbaum2002observational}.
\end{enumerate}

The first two approaches have the desirable property that, under a
specified deviation, one can often use existing theory to derive
an asymptotically normal (and sometimes efficient) estimator of the
causal effect. However, they are arguably more difficult to interpret
than Rosenbaum's sensitivity model because it is impossible to exhaust all possible
deviations in this way. Often, one considers just a few functional forms of
the deviation and hopes the results of the sensitivity analysis can be
extended to ``similar'' functional forms \citep[see
e.g.][]{brumback2004sensitivity}.

Our marginal sensitivity model can
be regarded as a hybrid of the selection model and Rosenbaum's
approach, in that we consider a range of possible differences between
$A|\bm X, Y(a)$ and $A | \bm X$. This model was considered first
introduced by \citet{tan2006distributional}, who noticed that
the range of IPW point estimates can be computed by linear
programming. However, \citet{tan2006distributional} did not consider
sampling variation of the bounds, thus his method has limited applicability in practice.

\subsection{Comparison with Rosenbaum's sensitivity analysis}
\label{sec:comp-with-rosenb}

\citet{rosenbaum1987sensitivity} proposed to quantify the degree
of violation of the MAR/NUC assumption based on the largest odds ratio
of $e_0(\bm x,y_1)$ and $e_0(\bm x,y_2)$:

\begin{defn}[Rosenbaum's Sensitivity Models] \label{def:rosenbaum-sens-model} Fix a parameter a $\Gamma \geq 1$ which will quantify the degree of violation from the MAR assumption.
  \begin{enumerate}
  \item  For the missing data problem, assume $e(\cdot, \cdot ) \in \mathcal{R}(\Gamma) $, where
    \begin{equation} \label{eq:rosenbaum-sens-model}
      \mathcal{R}(\Gamma) = \Big\{ e(\cdot, \cdot ):\,\frac{1}{\Gamma} \le \mathrm{OR}(e(\bm x,y_1), e(\bm x,y_2)) \leq \Gamma, \text{ for all }\bm x \in\mathscr X,y_1,y_2 \in \mathbb{R}\Big\}.
    \end{equation}
  \item For the observational studies problem, assume $e_a(\cdot, \cdot )  \in \mathcal{R}(\Gamma)$, for $a = 0, 1$.
  \end{enumerate}
\end{defn}

The proof of \Cref{ppn:complete-data} indicates that not all $e_0(\bm
X,Y)$ are compatible with the observed data. To compare the marginal
sensitivity model $\mathcal{E}$ with Rosenbaum's model $\mathcal{R}$,
we introduce the following concept:
\begin{defn}[Compatibility of Sensitivity Model]
  A sensitivity model $e_0(\bm x, y) = \P_0(A=1|\bm X
  = \bm x, Y = y)$ is called \emph{compatible} if the RHS of
  \eqref{eq:identifiable-proof} integrates to $1$, for all $\bm x \in \sX$, that is, $e_0(\cdot, \cdot )  \in \mathcal C$, where
  \[
    \mathcal{C} = \bigg\{ e(\cdot, \cdot): \int \mathrm{OR}(e_0(\bm x), e(\bm
    x, y))  \diff \P_0(y|A=1,\bm X = \bm x) = 1, ~\text{for all}~\bm x \in \sX \bigg\}.
  \]
\end{defn}
Then Rosenbaum's sensitivity model and the marginal sensitivity model are related in the following way:
\begin{ppn} \label{ppn:relation-rosenmbaum-our}For any $\Lambda \geq 1$,
  $\mathcal{E}(\sqrt{\Lambda}) \subseteq \mathcal{R}(\Lambda)$ and $\mathcal{R}(\Lambda) \cap \mathcal{C} \subseteq \mathcal{E}(\Lambda)\cap \mathcal{C}$.
\end{ppn}

As mentioned in the Introduction, Rosenbaum and his coauthors obtained point estimate and confidence interval of the causal effect under the collection of sensitivity models $\mathcal{R}(\Gamma)$. To this end, it is often assumed that the causal effect is additive and constant across
the individuals, that is, $Y_i(1) - Y_i(0) \equiv \Delta$, for all $i
\in [n]$. Then to determine if an effect $\Delta$ should
be included in the $(1-\alpha)$-confidence interval, one just needs to
test at level $\alpha$ the Fisher null $H_{0}:\,Y_i(0) = Y_i(1)$ for all $i \in [n]$,
using $Y - \Delta A$ as the outcome \citep{hodges1963estimates} and under
Rosenbaum's sensitivity model. We refer the reader to \citet[Chapter
4]{rosenbaum2002observational} for an overview of this
approach. Our approach is different from existing methods targeting
Rosenbaum's sensitivity model in many ways, sometimes markedly:
\begin{itemize}
\item {\it Population}: Most if not all existing methods for Rosenbaum's
  model treat the observed samples as the population,
  whereas we treat the observations as i.i.d.\ samples from a much
  larger super-population.
\item {\it Design}: Existing methods usually require the data are paired or
  grouped. Statistical theory assumes the matching is
  \emph{exact}, which is usually not strictly enforced in
  practice. Our approach is based on the IPW estimator and does not
  require exact matching.
\item {\it Sensitivity Model}: 
  We consider a different but closely related
  sensitivity model. Rosenbaum's sensitivity model is most natural for
  matched designs, whereas the marginal model is most natural when using IPW
  estimators. We also consider a parametric extension of the
  marginal sensitivity model.
\item {\it Statistical Inference}: Most existing methods are based on
  randomization tests of Fisher's sharp null hypothesis, utilizing the
  randomness in treatment assignment. Our approach takes a point
  estimation perspective by trying to estimate the average treatment
  effect directly. The distinction can be best understood by comparing
  to the distinction between hypothesis testing and point estimation,
  or in \citet{ding2017paradox}'s terminology, the subtle difference
  between Neyman's null (the \emph{average} causal effect is zero) and
  Fisher's null (the individual causal effects are \emph{all} zero).
\item {\it Effect Heterogeneity}: Constructing confidence intervals under
  Rosenbaum's sensitivity model usually require the causal effect is
  homogeneous, apart from \citet{rosenbaum2002attributing} who considered the
  ``attributable effect'' of a treatment. Some very recent advancements
  aim to remove this requirement in randomization
  inference. \citet{fogarty2017randomization} and
  \cite{fogarty2017sensitivity} considered estimating the sample ATE in
  observational studies with a matched pairs design. Our
  approach inherently allows the causal effect to be heterogeneous.
\item {\it Applicability to Missing Data Problems}: Our approach can be
  easily applied to missing data problems.
\end{itemize}

\subsection{Partially Identified Parameter}
\label{sec:part-ident-param}

Our framework is also related to a literature in econometrics on
partially identified parameters
\citep{imbens2004confidence,vansteelandt2006ignorance,chernozhukov2007estimation,aronow2012interval,miratrix2017shape}. See
\citet{richardson2014nonparametric} for a recent review. The mean response $\mu$ or the ATE $\Delta$ can be
regarded as partially identified under the marginal sensitivity
model. In fact, we have adopted the terminology ``partially
identified region'' for the set $\{\mu^{(h)}: e^{(h)} \in \mathcal
E_{\bm \beta_0}(\Lambda)\}$.

The main distinction is that existing methods in this literature
usually require estimates of the boundaries of the partially identified
region with known asymptotic distributions. In \Cref{sec:union-method}
we have shown that this is inherently difficult for sensitivity
analysis. Our work opens the door for inference of partially
identified parameters when it is difficult to analyze the asymptotic
behavior of the boundary estimates.


\section{Numerical Examples}
\label{sec:numer-illustr}

\subsection{Simulation study}
\label{sec:simulation-study}

\blue{
  We evaluate the performance of the proposed percentile bootstrap
  procedure with a simulation study for the missing data
  problem. Because obtaining the true partially identified interval
  generally requires solving two functional optimization problems, we
  restrict the covariate $X$ to be univariate and have discrete support
  and the response $Y$ to be binary. More specifically, we sample $n =
  200$ i.i.d.\ data points with $X$ uniformly
  distributed on $\{-2.5, 1.28, 0.54, 0.16,\allowbreak 0.02, 0, 0.02,
  0.16, 0.54, 1.28, 2.5\}$ and $Y$ generated by
  $\text{logit}\,\mathbb{P}(Y=1|X) = - \beta_Y X$. The missingness
  indicator $A$ for each data point is then generated by
  $\text{logit}\,e_0(X,Y) = \text{logit}\,e_0(X) = \text{logit}\,\mathbb{P}(A=1|X) =
  \beta_A X + 0.1 X^2$. The parameters $\beta_A$ and $\beta_Y$ are
  chosen to be $0.5$ or $1.5$ in the simulation (thus 4 settings in
  total). In the settings with $\beta_A = 1.5$, the selection
  probability can be very close to $0$ for some data points, a situation
  where the IPW estimator is known to be unstable.

  Six values of the sensitivity parameter $\lambda$ are considered:
  $\lambda = 0,0.1,0.2,0.5,1,2$. In each setting, we compute the
  partially identified interval for $\mathbb{E}[Y]$,
  $\{\mu^{(h)}:e^{(h)}(x,y) \in \mathcal{E}_{\beta_0}(\Lambda)\}$, by solving
  a linear programming problem on the population level:
  \begin{align*}
    \mathrm{minimize~or~maximize}~\frac{\sum_x \mathbb{E}[Y|X=x] \mathbb{P}(A=1|X=x) /
    e^{(h)}(x,y)}{\sum_x \sum_{y=0}^1 \mathbb{P}(A=1|X=x) / e^{(h)}(x,y)}
  \end{align*}
  over $e^{(h)}(x,y) \in \mathcal{E}_{\beta_0}(\Lambda)$.
  Solving this population optimization problem
  is possible because $X$ and $Y$ are discrete. We use the
  percentile bootstrap procedure in
  \Cref{sec:conf-interv-under} to construct the interval of point
  estimates and the confidence interval (nominal level is set to
  $90\%$). This is repeated for $1000$ times to obtain the
  non-coverage rate of the confidence intervals. We also report the
  median interval for the point estimates and the median confidence
  interval by taking the sample median of the corresponding extrema in
  the $1000$ repetitions.

  Notice that when applying our procedure, we only include the linear
  term in modeling $e(X)$, i.e.\ $\mathrm{logit}\,e_{\beta}(X) =
  X\beta$, thus the missingness model is misspecified. Using a
  large-sample numerical approximation, we find the parametric model
  $e_{\beta_0}(X)$ that is closest to $e_0(X)$ in Kullback-Leibler
  divergence. The largest
  difference between $\mathrm{logit}\,e_{\beta_0}(x)$ and
  $\mathrm{logit}\,e(x)$ is smaller than $0.65$ in all 4
  settings, i.e.\ $e_0(x) \in \mathcal{E}_{\beta_0}(e^{0.65})$. However
  for most $x$ the difference is much smaller and close to $0$.

  The simulation results are reported in \Cref{tab:simulation}. The
  confidence intervals constructed by the percentile bootstrap have
  desired coverage (90\%) when $\beta_A = 0.5$ and do not appear to be
  conservative, even though we used the generalized minimax inequality
  in \Cref{lem:minimax} to prove the main result. However, when $\beta_A
  = 1.5$, the confidence intervals are anti-conservative and the
  non-coverage rate becomes larger as $\lambda$ increases. This is most
  likely due to the well known phenomenon that the IPW estimators are
  unstable when the selection probability is close to $0$
  \citep{kang2007demystifying} and the relative small sample size being
  used ($n = 200$). The instability problem is exacerbated as the
  sensitivity value increases.

  Regardless of the value of $\beta_A$, the median interval of point
  estimates is always very close to the true partially identified
  interval. This shows that the range of SIPW estimates constructed by
  solving the optimization problem \eqref{eq:lp} is unbiased in the
  simulation.

  A final point we want to point out is that the bias due to
  misspecifying $e_0(x)$ is small. Due to symmetry the true mean
  response is $\mu_0 = \mathbb{E}_0[Y] = 0.5$. The large-sample limit of
  the SIPW estimator can be found on the rows with $\lambda = 0$ under
  column ``Part.\ iden.\ int.''. The one that is most different from
  $\mu_0 = 0.5$ is $0.528$ for $\beta_A = \beta_Y = 1.5$. The partially
  identified interval for $\lambda = 0.65$ would cover the true mean
  response $\mu_0 = 0.5$ because $e_0(x) \in
  \mathcal{E}_{\beta_0}(e^{0.65})$. However, using such a large value of
  $\lambda$ is not necessary in all four settings. In three settings
  using $\lambda = 0.1$ (and in the other setting using $\lambda = 0.2$)
  would make the partially identified interval cover $\mu_0 =
  0.5$. This shows that the $L_{\infty}$-sensitivity model we use may be
  conservative for modeling misspecification bias.
}

\small
\renewcommand*{\arraystretch}{1.2}
\begin{table} \color{blue}
  \caption{\color{blue} \small{Simulation results for estimating the mean response in some
    missing data problems. The first four columns index settings of the
    simulation study: $\beta_A$ ($\beta_Y$) is the strength of
    association between $X$ and $A$ ($Y$); $\lambda$ is the sensitivity
    parameter and $\Lambda = e^{\lambda}$. The next five columns are the
    results: non-coverage rate of the confidence intervals constructed by
    our methods (desired non-coverage rate is $10\%$); partially identified interval
    of the mean response; median interval for the point estimates; median
    confidence interval.}}
  \label{tab:simulation}
  \begin{tabular}{llll|cccc}
    \hline
    $\beta_A$ & $\beta_Y$ & $\lambda$ & $\Lambda$ & Non-coverage &                                                                 Part.\
                                                                   iden.\ int.
    & Point est.\ int.\ & Conf.\ int. \\
    \hline
    0.5 & 0.5 & 0  & 1 & $0.103$ & $[0.506, 0.506]$ & $[0.506, 0.506]$ & $[0.422, 0.584]$ \\
              &  & 0.1  & 1.11 & $0.094$ & $[0.482, 0.530]$ & $[0.481, 0.529]$ & $[0.404, 0.613]$ \\
              &  & 0.2  & 1.22 & $0.115$ & $[0.458, 0.553]$ & $[0.457, 0.552]$ & $[0.374, 0.631]$ \\
              &  & 0.5  & 1.65 & $0.092$ & $[0.388, 0.622]$ & $[0.389, 0.623]$ & $[0.310, 0.698]$ \\
              &  & 1  & 2.72 & $0.127$ & $[0.282, 0.728]$ & $[0.283, 0.729]$ & $[0.214, 0.789]$ \\
              &  & 2  & 7.39 & $0.115$ & $[0.130, 0.876]$ & $[0.129, 0.876]$ & $[0.092, 0.911]$ \\
              & 1.5 & 0  & 1 & $0.109$ & $[0.510, 0.510]$ & $[0.510, 0.510]$ & $[0.436, 0.584]$ \\
              &  & 0.1  & 1.11 & $0.106$ & $[0.486, 0.533]$ & $[0.487, 0.534]$ & $[0.415, 0.609]$ \\
              &  & 0.2  & 1.22 & $0.092$ & $[0.462, 0.556]$ & $[0.462, 0.557]$ & $[0.392, 0.630]$ \\
              &  & 0.5  & 1.65 & $0.101$ & $[0.392, 0.627]$ & $[0.392, 0.626]$ & $[0.322, 0.696]$ \\
              &  & 1  & 2.72 & $0.110$ & $[0.286, 0.730]$ & $[0.286, 0.731]$ & $[0.226, 0.790]$ \\
              &  & 2  & 7.39 & $0.096$ & $[0.132, 0.878]$ & $[0.132, 0.878]$ & $[0.096, 0.911]$ \\
    1.5 & 0.5 & 0  & 1 & $0.135$ & $[0.515, 0.515]$ & $[0.516, 0.516]$ & $[0.380, 0.620]$ \\
              &  & 0.1  & 1.11 & $0.159$ & $[0.489, 0.541]$ & $[0.489, 0.541]$ & $[0.361, 0.642]$ \\
              &  & 0.2  & 1.22 & $0.170$ & $[0.463, 0.566]$ & $[0.465, 0.567]$ & $[0.343, 0.659]$ \\
              &  & 0.5  & 1.65 & $0.186$ & $[0.388, 0.641]$ & $[0.388, 0.641]$ & $[0.283, 0.725]$ \\
              &  & 1  & 2.72 & $0.270$ & $[0.274, 0.748]$ & $[0.276, 0.750]$ & $[0.194, 0.809]$ \\
              &  & 2  & 7.39 & $0.337$ & $[0.119, 0.892]$ & $[0.119, 0.892]$ & $[0.082, 0.921]$ \\
              & 1.5 & 0  & 1 & $0.188$ & $[0.528, 0.528]$ & $[0.525, 0.525]$ & $[0.393, 0.612]$ \\
              &  & 0.1  & 1.11 & $0.172$ & $[0.501, 0.552]$ & $[0.502, 0.553]$ & $[0.369, 0.639]$ \\
              &  & 0.2  & 1.22 & $0.200$ & $[0.476, 0.578]$ & $[0.476, 0.578]$ & $[0.350, 0.655]$ \\
              &  & 0.5  & 1.65 & $0.190$ & $[0.401, 0.650]$ & $[0.400, 0.650]$ & $[0.288, 0.724]$ \\
              &  & 1  & 2.72 & $0.231$ & $[0.286, 0.755]$ & $[0.284, 0.754]$ & $[0.201, 0.810]$ \\
              &  & 2  & 7.39 & $0.248$ & $[0.126, 0.893]$ & $[0.126, 0.894]$ & $[0.085, 0.925]$ \\
    \hline
  \end{tabular}
\end{table}

\normalsize

\subsection{Real data example}
\label{sec:real-data-example}

Finally we illustrate the methods proposed in
\Cref{sec:new-framework,sec:extension} by an observational study, in
which we are interested in estimating the causal effect of fish consumption on
the blood mercury level. We obtained $2512$ survey responses from the
National Health and Nutrition Examination
Survey (NHANES) 2013-2014 who were at least $18$ years old, answered
the questionnaire about seafood consumption, and had blood mercury
measured. Among these individuals, $1$
has missing education, $7$ have missing smoking, and
$175$ have missing income. We removed the individuals with missing
education or smoking and imputed the missing income using the median
income (we also added a binary indicator for missing income). Then we
defined high fish consumption as more than 12 servings of fish or
shellfish in the previous month, and low fish consumption as 0 or 1
servings of fish. In the end we were left with 234 treated individuals
(high consumption), $873$ controls (low consumption), and $8$
covariates: gender, age, income, whether income is missing, race,
education, ever smoked, and number of cigarettes smoked last
month. The outcome variable is $\log_2$ of total blood mercury
(in ug/L). This dataset was also analyzed by \citet{zhao2017cross} and is
publicly available in the \texttt{R} package \texttt{CrossScreening}
on CRAN.

We used the percentile bootstrap to conduct sensitivity analyses for
four estimators: $\hat{\Delta}$,
$\hat{\Delta}_{\mathrm{SAIPW}}$,
$\hat{\Delta}_{1}$, and
$\hat{\Delta}_{1, \mathrm{SAIPW}}$. The propensity score $\mathbb{P}(A=1|\bm X)$ is
estimated by a logistic regression and the outcome means
$\mathbb{E}[Y(1)|\bm X]$ and $\mathbb{E}[Y(0)|\bm X]$ are
estimated by linear regressions using all $8$ covariates. \blue{Using the
  entire dataset, the estimated propensity score ranges from $0.014$ to
  $0.794$. Since some propensity scores are close to $0$, results
  of an IPW estimator (in particular coverage properties of the
  confidence interval) should be interpreted cautiously following our
  simulation study in \ref{sec:simulation-study}.} We used $B =
1000$ bootstrap samples to obtain $90\%$ confidence intervals of
$\Delta$ and $\Delta_1$ under $\mathcal E_{\bm \beta_0}(\Lambda)$ for
$5$ values of $\lambda = \log \Lambda = 0, 0.5, 1, 2, 3$.

We compared the results with Rosenbaum's sensitivity analysis as
implemented in the \texttt{senmwCI} function (default options) in the
\texttt{R} package \texttt{sensitivitymw} \citep{rosenbaum2015two}. We
used the $234$ matched pairs created by \citet{zhao2017cross} as the
basis of the sensitivity analysis. As mentioned previously,
Rosenbaum's sensitivity analysis assumes constant treatment effect
(CTE) to construct confidence intervals.

\begin{figure}[t]
  \centering
  \includegraphics[width = 0.75\textwidth]{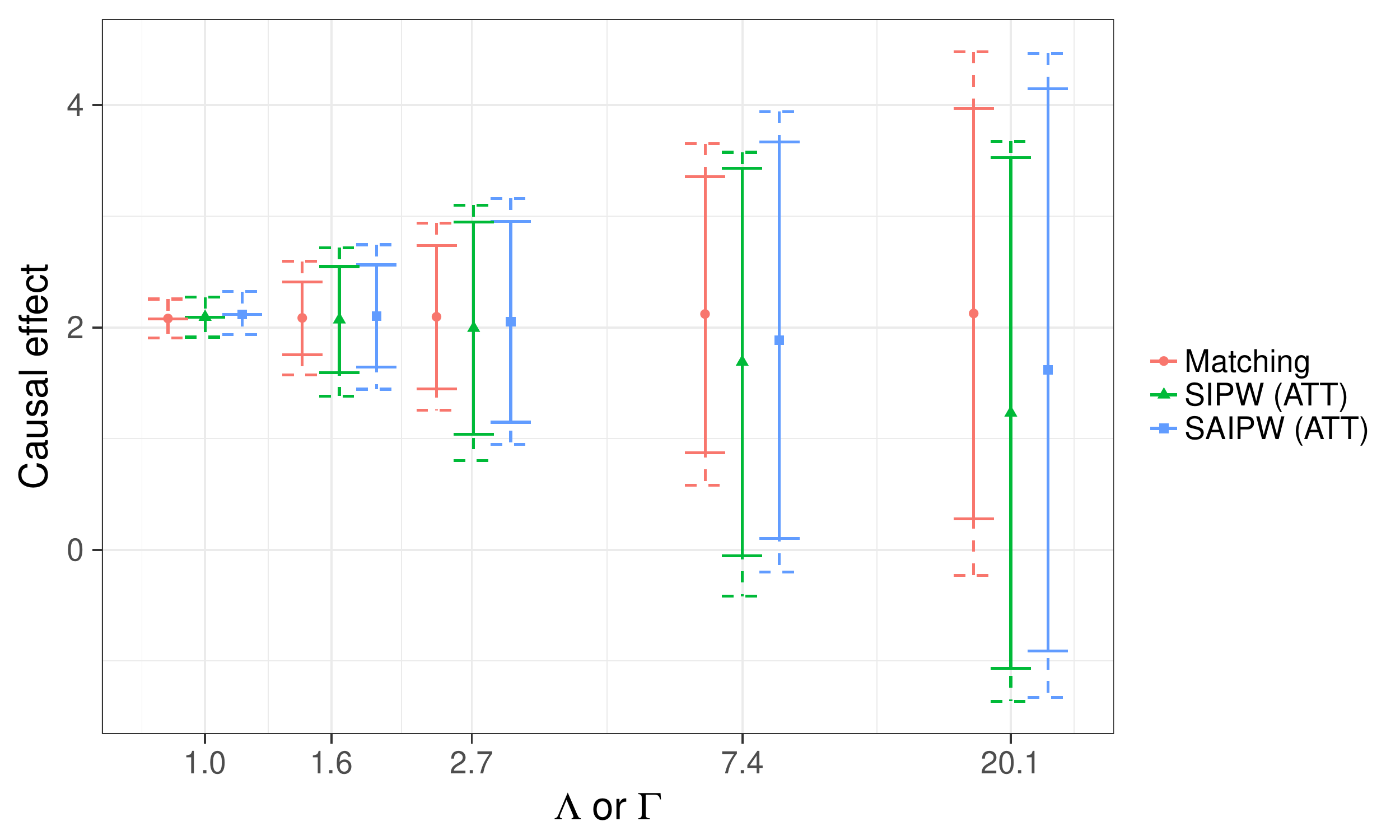}
  \caption{\small Graphical illustration of the sensitivity analysis
    results for three methods (matching, SIPW for ATT, and SAIPW for
    ATT). The solid error bars are the range of point estimates and
    the dashed error bars (together with the solid bars) are
    the confidence intervals. The circles/triangles/squares are the
    mid-points of the solid bars.}
  \label{fig:fish-mercury}
\end{figure}

The results of the five sensitivity analyses are reported in
\Cref{tab:fish-mercury}. Alternatively, one can report the results
by plotting the confidence intervals against $\Lambda$
(\Cref{fig:fish-mercury}). Overall, the confidence intervals constructed
by the percentile bootstrap were slightly wider than those constructed
by Rosenbaum's sensitivity analysis under the same $\Lambda$, while the
percentile bootstrap intervals under $\sqrt{\Lambda}$ were shorter than
Rosenbaum's under $\Lambda$. This observation is not surprising given
\Cref{ppn:relation-rosenmbaum-our}.
Augmentation by outcome
regressions (SAIPW estimators) helped to reduce the width of confidence
intervals when the estimand is ATE, but did not reduce the width when
the estimand is ATT. The IPW analyses suggested that the ATE/ATT is
significantly positive for at least $\Lambda = 2.72$, while the matching analysis
found the effect is significantly positive for at least $\Gamma =
7.39$. \blue{This means that, in order for the ATE/ATT to be non-significant, the estimated propensity score must be
  quite different from the actual propensity score given all
  confounders. We consider this as a fairly strong evidence that the
  qualitative conclusion ``consuming fish increases blood mercury
  level'' is somewhat insensitive to unmeasured confounding.}

Lastly we want to comment on the computational costs reported in the
last column of \Cref{tab:fish-mercury}. The IPW analyses are slower
than the matching analyses (assuming the matches are given), but the
running time is quite acceptable given we have over a thousand
observations. The reason for the apparent advantage of matching is
that we do not include the time for generating the matches. The
matching analysis uses analytic
approximations to conduct sensitivity analysis, hence it is faster
than the IPW analyses which use the bootstrap. Since the time
complexity of the IPW analyses
scales almost linearly with the sample size, we expect the running times for
larger studies will still be acceptable. \\

\begin{table}[t]
  \caption{\small{Results of different sensitivity analyses under the
      collection of parametric sensitivity models $\mathcal{E}_{\bm
        \beta_0}(\Lambda)$. Five methods were considered: the SIPW
      and SAIPW estimators of ATE and ATT as described in
      \Cref{sec:new-framework,sec:extension}, and Rosenbaum's sensitivity
      analysis based on matched pairs. Eight sensitivity parameters
      were used: $\lambda = \log \Lambda$ or $\gamma = \log \Gamma$ $= 0, 0.5, 1, 2, 3$. The
      running time of these five methods were also reported. For the first four
      methods, we used $B=1000$ bootstrap samples and used $3$ cores in
      parallel to obtain the confidence intervals. For matching, we
      do not include the computational cost of generating the matches.}}
  \label{tab:fish-mercury}
  \scriptsize
  \begin{tabular}{|ll|cc|cc|cc|}
    \hline
    \rule{0pt}{3ex}
    Estimand & Method & Point & 90\% CI & Point & 90\% CI & Point &
                                                                    \multicolumn{1}{c|}{90\% CI} \\
    \hline
             & & \multicolumn{2}{c|}{$\Lambda = e^0 = 1$} & \multicolumn{2}{c|}{$\Lambda =
                                                            e^{0.5} = 1.65$} & \multicolumn{2}{c|}{$\Lambda = e^{1} = 2.72$} \\
    ATE & SIPW  & (1.86,1.86) & (1.63,2.06) & (1.33,2.37) & (1.11,2.55) & (0.83,2.84) & (0.61,2.99) \\
             & SAIPW  & (1.80,1.80) & (1.55,2.04) & (1.34,2.28) & (1.16,2.53) & (0.89,2.76) & (0.73,2.99) \\
    ATT & SIPW  & (2.09,2.09) & (1.91,2.29) & (1.59,2.55) & (1.38,2.72) & (1.04,2.95) & (0.80,3.12) \\
             & SAIPW  & (2.12,2.12) & (1.93,2.31) & (1.64,2.56) & (1.45,2.74) &
                                                                                (1.15,2.95) & (0.95,3.16) \\
    \hline
             & & \multicolumn{2}{c|}{$\Gamma = e^0 = 1$} & \multicolumn{2}{c|}{$\Gamma =
                                                           e^{0.5} = 1.65$} & \multicolumn{2}{c|}{$\Gamma = e^{1} = 2.72$} \\
    Constant & Matching  & (2.08,2.08) & (1.90,2.25) & (1.75,2.41) & (1.57,2.59) & (1.45,2.74) & (1.25,2.94) \\
    \hline
    \hline
    \rule{0pt}{3ex} Estimand & Method & Point & 90\% CI & Point & 90\% CI &
                                                                            \multicolumn{2}{c|}{Running time (seconds)} \\
    \hline
             & & \multicolumn{2}{c|}{$\Lambda = e^2 = 7.39$} & \multicolumn{2}{c|}{$\Lambda =
                                                               e^3 = 20.09$} & & \\
    ATE & SIPW  & (-0.10,3.78) & (-0.30,4.01) & (-0.91,4.78) &
                                                               (-1.15,4.99) &  \multicolumn{2}{c|}{32.8} \\
             & SAIPW  & (0.12,3.61) & (-0.02,3.83) & (-0.55,4.31) & (-0.76,4.55) &
                                                                                   \multicolumn{2}{c|}{47.7}\\
    ATT & SIPW  & (-0.05,3.43) & (-0.43,3.58) & (-1.07,3.53) &
                                                               (-1.36,3.68) &  \multicolumn{2}{c|}{18.0} \\
             & SAIPW  & (0.10,3.67) & (-0.20,3.92) & (-0.91,4.15) & (-1.36,4.47) &
                                                                                   \multicolumn{2}{c|}{29.9}\\
    \hline
             & & \multicolumn{2}{c|}{$\Gamma = e^0 = 1$} & \multicolumn{2}{c|}{$\Gamma =
                                                           e^{0.5} = 1.65$} & & \\
    Constant & Matching  & (0.87,3.36) & (0.58,3.65) & (0.28,3.97) &
                                                                     (-0.23,4.48) &  \multicolumn{2}{c|}{1.8} \\
    \hline
  \end{tabular}
\end{table}

%

{\noindent\textbf{Acknowledgement}: The authors thank Colin Fogarty for pointing out the difference between Rosenbaum's  sensitivity model and the marginal sensitivity model.}

\small
\bibliographystyle{plainnat}
\bibliography{ref}

\normalsize

\appendix
\section{Proofs}
\label{sec:proofs-main}

Here we prove \Cref{ppn:union-ci} and \Cref{lem:minimax}. Additional proofs can be found in the supplementary file.

\subsection{Proof of \Cref{ppn:union-ci}}
\label{sec:proof-crefppn:-ci}

By definition, under the sensitivity model $\cH(\gamma)$, the true data generating distribution $F_0$ satisfies $h_0(\bm x, y) \in \mathcal{H}_{\gamma}$.
 This implies that
  \begin{equation*} 
    \P_0(\mu \in [L,U]) = \mathbb{P}_0\big(\mu^{(h_0)} \in [L,U]\big) \ge
    \mathbb{P}_0\big(\mu^{(h_0)} \in [L^{(h_0)}, U^{(h_0)}]\big).
  \end{equation*}
  The last inequality is true because $[L,U] \supseteq [L^{(h_0)},U^{(h_0)}]$. Now, taking
  limit on both sides gives
\[
\liminf_{n \rightarrow \infty}  \P_0(\mu
  \in [L,U])\geq \liminf_{n \rightarrow \infty}
  \mathbb{P}_0\big(\mu^{(h_0)} \in [L^{(h_0)}, U^{(h_0)}]\big) \ge
  1-\alpha,
\]
since $[L^{(h_0)}, U^{(h_0)}]$ is an asymptotically $(1-\alpha)$-confidence interval for $\mu^{(h_0)}$.

  For the second part of the proposition, let $h_{\mathrm{min}} = \argmin_{h \in \mathcal{H}(\gamma)}
  \mu^{(h)}$. By applying assumption \eqref{eq:extension-alpha-prime} to the intervals $[L^{(h)},\infty)$, for $h \in \mathcal{H}(\gamma)$, gives $\limsup_{n \to \infty} \P_0\big(\mu^{(h_{\mathrm{min}})} < L\big) \le \alpha'$. Similarly, defining $h_{\mathrm{max}} = \argmax_{h \in \mathcal{H}(\gamma)}
  \mu^{(h)}$, we have $\limsup_{n \to \infty} \P_0\big(\mu^{(h_{\mathrm{max}})} > U\big) \le \alpha - \alpha'$. Then using the union bound,
\begin{align*}
\limsup_{n \to \infty} \P_0 & \left(\{\mu^{(h)}: e^{(h)} \in \mathcal E_{\bm \beta_0}(\Gamma)\} \not\subseteq [L,U]\right) \\
& \le \limsup_{n \to \infty} \P_0\left(\left[\mu^{(h_{\mathrm{min}})},\mu^{(h_{\mathrm{max}})}\right] \not \subseteq [L, U]\right) \\
& \le \limsup_{n \to \infty} \P_0\left(\mu^{(h_{\mathrm{min}})} < L\right) +
\limsup_{n \to \infty} \P_0\left(\mu^{(h_{\mathrm{max}})} > U\right) \\
& \le \alpha,
\end{align*}
completing the proof.

\subsection{Proof of \Cref{lem:minimax}}
\label{sec:proof-crefl}

For $1\leq b \leq N$, where $N=n^n$ is the total number of possible bootstrap resamples, denote by $\hat{\hat{\mu}}^{(h)}_{b}$ the estimate \eqref{eq:sipw} in the $b$-th resample. Therefore, for every $h \in \mathcal{H}(\gamma)$,
  \begin{equation} \label{eq:minimax-proof-1}
    \hat{\hat{\mu}}^{(h)}_{b} \ge \inf_{h \in \mathcal{H}(\gamma)} \hat{\hat{\mu}}^{(h)}_{b},~\text{ for all } 1 \le b \le N.
  \end{equation}
  Note that both sides of \eqref{eq:minimax-proof-1} above are sequences indexed by
  $b \in [N]$, and since the inequality is true entry by entry, it is also true for
  any order statistic of the sequences, namely for any $0 < \alpha < 1$,
  \begin{equation} \label{eq:minimax-proof-2}
    Q_{\frac{\alpha}{2}}\bigg(\Big(\hat{\hat{\mu}}^{(h)}_{b}\Big)_{b \in [N]} \bigg) \ge Q_{\frac{\alpha}{2}}\bigg(\Big(\inf_{h \in \mathcal{H}(\gamma)} \hat{\hat{\mu}}^{(h)}_{b}\Big)_{b \in [N]}\bigg)=L,
  \end{equation}
  by definition \eqref{eq:pb-ci}. Since \eqref{eq:minimax-proof-2} is true for any $h \in
  \mathcal{H}(\gamma)$, taking infimum on the LHS above gives
  \[ \inf_{h \in \mathcal{H}(\gamma)} L^{(h)} = \inf_{h \in
    \mathcal{H}(\gamma)} Q_{\frac{\alpha}{2}}\bigg(\Big(\hat{\hat{\mu}}^{(h)}_{b}\Big)_{b \in [N]} \bigg) \ge L.\]
  The lower bound on $U$ can be proved similarly.

\subsection{Proof of \Cref{ppn:lfp}}
\label{sec:proof-crefppn:lfp}

We prove the first claim by contradiction. Suppose there exists two indices $s_1 < s_2$ such that $A_{s_1} = A_{s_2} = 1$, $Y_{s_1} > Y_{s_2}$ but $z_{s_1} < z_{s_2}$. Then consider the following perturbation,
  \[
  z'_i =
  \begin{cases}
    z_i, & s  \ne s_1, s_2, \\
    z_{s_1} + \varepsilon e^{\hat{g}(\bm X_{s_1})}, & i= s_1, \\
    z_{s_2} - \varepsilon e^{\hat{g}(\bm X_{s_2})}, & s = s_2. \\
  \end{cases}
  \]
  When $\varepsilon > 0$ is sufficiently small, $(z_i')_{i=1}^m$ is still feasible but the objective becomes larger, which contradicts the assumption that $(z_i)_{i=1}^m$ is the maximizer.

  Next, we prove the second claim. It is well known that if a linear programming is feasible and bounded, then there is at least one vertex (also called the basic feasible) solution \citep{dantzig1951maximization}. Notice that in \eqref{eq:lp}, there are $m+1$ optimization variables and $1$ equality constraint. It is also easy to verify that $t = 0$ is not feasible. Therefore, among the $2m$ inequality constraints, $ \frac{1}{\Gamma} t \le \bar z_i \le \Gamma t$, $1 \leq s \leq m$, there exists a solution of \eqref{eq:lp} such that
  $m$ equalities hold. This implies that $z_i=\frac{1}{t}\bar z_i$ is either
  $\Gamma$ or $ \frac{1}{\Gamma}$, for all $1 \leq s \leq m$. Then, using the first
  part of the proposition, there exists a $M$
  such that $z_i = \Gamma$, if $Y_i \ge M$, and $z_i = \frac{1}{\Gamma}$, if
  $Y_i < M$.

\subsection{Proof of \Cref{ppn:relation-rosenmbaum-our}}
\label{sec:proof-crefppn:r-rose}

To begin with suppose that $e \in \mathcal{E}(\sqrt{\Gamma})$. Then, for all $\bm x \in \sX$, and $y_1, y_2 \in \R$,
\begin{align*}
|\log \mathrm{OR}(e(\bm x,y_1), e(\bm x,y_2))| & \le |\log \mathrm{OR}(e(\bm x,y_1), e_0(\bm x))| + |\log \mathrm{OR}(e_0(x), e(\bm x,y_2))| \nonumber \\
& \le \log \Gamma,
\end{align*}
which implies $e \in \mathcal{R}(\Gamma)$.

Now, suppose the function $e \in \mathcal{R}(\Gamma) \cap \mathcal{C}$. By \eqref{eq:rosenbaum-sens-model},
\[
\frac{1}{\Gamma} \le \mathrm{OR}\left(\inf_{y \in \R} e(\bm x, y),\, \sup_{y \in \R} e(\bm
x, y)\right) \le \Gamma,~\text{for all}~\bm x \in \sX.
\]
  Notice that $\inf_y e(\bm x, y) \le e_0(\bm x) \le \sup_y e(\bm x, y)$,
  because $e(\bm x,y)$ marginalizes to $e_0(\bm x)$ and $e(\bm x,y)$ is
  compatible. Thus, $\mathrm{OR}(e(\bm x, y), e_0(\bm x))$ must be
  between $1/\Gamma$ and $\Gamma$, which implies $ e \in \mathcal{E}(\Gamma)\cap \mathcal{C}$.


\section{Unidentifiability of $e_0(\textit{\textbf{x}},y)$}
\label{sec:proof-crefppn:c-data}

\begin{ppn} \label{ppn:complete-data} In the missing data problem and
  assuming \Cref{assump:overlap} holds, $e_0(\bm x ,y) = \mathbb{P}_0(A=1|\bm X=\bm x,Y=y)$ is not identifiable from the data. 
\end{ppn}

\noindent{\it Proof}: The complete data density can be factorized as
  \[\P_0(A,\bm X,Y) = \P_0(A,\bm X) \cdot \P_0(Y|A,\bm X).\]
  The first term in the RHS above is identifiable from the data, because $A$ and $\bm X$ are always observed. For the second term, note that $\P_0(Y|A=1,\bm X)$ is identifiable because $Y$ is observed if $A=1$. For $\P_0(Y|A=0,\bm X)$, using Bayes rule, we get
  \begin{equation*}
    \begin{split}
      &\P_0(Y|A=0,\bm X) \\
      =& \frac{\P_0(A=0|Y,\bm X) \cdot
        \P_0(Y|\bm X)}{\P_0(A=0|\bm X)} \\
      =&
      \frac{\P_0(A=0|Y,\bm X) \cdot \left\{ \P_0(Y|A=0,\bm X)\P_0(A=0|\bm X) + \P_0(Y|A=1,\bm X)\P_0(A=1|\bm X)\right\}}{\P_0(A=0|\bm X)}.
    \end{split}
  \end{equation*}
  Notice that if the denominator $\P_0(A = 0|\bm X = \bm x) = 0$, then
  we don't need to consider the conditional distribution of $Y$. By simple
  algebra, we have
  \begin{equation}
    \label{eq:identifiable-proof}
    \P_0(Y|A=0,\bm X) = \mathrm{OR}(e_0(\bm X), e_0(\bm
    X, Y)) \cdot \P_0(Y|A=1,\bm X),
  \end{equation}
  where $\mathrm{OR}(p_1,p_2) := [p_1/(1-p_1)]/[p_2/(1-p_2)]$ is the
    {\it odds ratio} of $p_1, p_2 \in (0, 1)$.
  The only terms in the above equation that are not directly identifiable
  from the data are $\P_0(Y|A=0,\bm X)$ and $e_0(\bm X, Y)$. We can
  arbitrarily specify $e_0(\bm x, y)$ as long as the
  RHS of
  \eqref{eq:identifiable-proof} integrates to $1$. This gives us
  $\P_0(Y|A=0,\bm X)$ and hence the complete data distribution.
   It is easy to see that
  different choices of $e_0(\bm x, y)$ can generate the
  same density for the observed data $(A, \bm X, AY)$ but different
  densities for the complete data $(A, \bm X, Y)$. \hfill $\Box$

\section{Proof of Theorem \ref{thm:valid-pbz}}
\label{sec:proofs}

To begin with, define  $\kappa^{(h)}:=\E_0 \left[ A \left(1 + e^{h(\bm X, Y) - \mathrm{logit}(\P_{\bm \beta_0}(A=1|\bm X))}  \right)\right]$.   Recall from \eqref{eq:muh}, that $$\mu^{(h)} = \frac{1}{\kappa^{(h)}} \E_0 \left[ AY \left(1 + e^{h(\bm X, Y) - \mathrm{logit}(\P_{\bm \beta_0}(A=1|\bm X))}  \right)\right],$$ for $h \in \cH(\gamma)$. We begin by showing that the estimates of the  parameters $(\mu^{(h)}, \kappa^{(h)} ,\bm \beta_0')' $ can be derived using the framework of $Z$-estimation.

\subsection{The $Z$-Estimation Framework} Given a vector $\bm \upsilon=(\nu, \kappa ,\bm \beta')' \in \Theta \subset \R \times \R_+ \times \R^d$, where $\Theta$ is the compact parameter space, define the function $Q: \{0, 1\} \times \R \times \R^d \rightarrow \R^{d+2}$ as follows: For $\bm t=(a, \bm x', y)' \in \{0, 1\} \times \R \times \R^d$,
\begin{align}\label{eq:Q}
  Q(\bm t|\bm \upsilon)= \left(
    \begin{array}{c}
      Q_1(\bm t|\bm \upsilon)   \\
      Q_2(\bm t|\bm \upsilon)    \\
      Q_3(\bm t|\bm \upsilon)
    \end{array}
  \right):=
  \left(
    \begin{array}{c}
      \left(a-\frac{e^{\bm \beta' \bm x}}{1+e^{\bm \beta' \bm x}}\right) \bm x     \\
      \kappa - a \left( 1+ e^{h(\bm x,  y)-\bm \beta' \bm x}\right)     \\
      \kappa  \nu - a y \left( 1+ e^{h(\bm x,  y)-\bm \beta' \bm x}\right)
    \end{array}
  \right).
\end{align}
Next, define $\Phi(\bm \upsilon)=\int Q(\bm t|\bm \upsilon) \mathrm d\P_0(\bm t)$, where  $\bm T=(A, \bm X', AY)' \sim \P_0$, the true distribution generating the data. Note that $\Phi(\bm \upsilon_0)=0$, where  $\bm \upsilon_0=(\mu^{(h)}, \kappa^{(h)}, \bm \beta_0')$ is the true parameter value. The $Z$-estimates $\hat {\bm \upsilon}=(\hat \mu^{(h)}, \hat \kappa^{(h)} , \hat{\bm \beta}')' $ are obtained by solving the equations
\begin{align}\label{eq:phi}
  \Phi_n(\hat {\bm \upsilon}):=\frac{1}{n}\sum_{i=1}^n Q(\bm T_i|\hat {\bm \upsilon})=
  \left(
    \begin{array}{c}
      \frac{1}{n}\sum_{i=1}^n \left(A_i-\frac{e^{\hat{\bm \beta}' \bm X_i}}{1+e^{\hat{\bm \beta}' \bm X_i}}\right) \bm X_i     \\
      \hat \kappa^{(h)} - \frac{1}{n}\sum_{i=1}^n A_i \left( 1+ e^{h(\bm X_i,  Y_i)-\hat{\bm \beta}' \bm X_i}\right)     \\
      \hat \kappa^{(h)}  \hat \mu^{(h)} - \frac{1}{n}\sum_{i=1}^n A_i Y_i \left( 1+ e^{h(\bm X_i,  Y_i)-\hat{\bm \beta}' \bm X_i}\right)
    \end{array}
  \right) = 0.
\end{align}
It is easy to see that the $Z$ estimate $\hat \mu^{(h)}$ is exactly the SIPW estimate \eqref{eq:sipw} for $\mu^{(h)}$ and $\hat {\bm \beta}$ is the MLE of $\bm \beta$ for the logistic regression model.

To formally define the bootstrap estimates, let $\P_n$ be the empirical measure of the sample $\bm T_1, \bm T_2, \ldots, \bm T_n$, where $\bm T_i=(A_i, \bm X_i', A_iY_i)$, and $\hat {\bm T_1}, \hat {\bm T_2}, \ldots, \hat {\bm T_n}$ be i.i.d. samples from the empirical measure. The bootstrap empirical distribution and the bootstrap empirical process are
$$\hat \P_n=\frac{1}{n}\sum_{i=1}^n \delta_{\hat{\bm T_i}} \quad \text{and} \quad \hat \G_n=\sqrt n(\hat \P_n-\P_n),$$
respectively. Noting that $\hat \Phi_n( \bm \upsilon) =\int Q(\bm t| \bm \upsilon) \mathrm d \hat \P_n(\bm t)$, the bootstrap $Z$-estimates $\hat{\hat {\bm \upsilon}}$ are obtained from the equations:
\begin{align}\label{eq:phihat}
  \hat \Phi_n(\hat{\hat {\bm \upsilon}}):=\frac{1}{n}\sum_{i=1}^n Q(\hat{\bm T_i}|\hat{\hat {\bm \upsilon}})=
  \left(
    \begin{array}{c}
      \frac{1}{n}\sum_{i=1}^n \left(\hat A_i-\frac{e^{\hat{\hat{\bm \beta}}'\hat{\bm X}_i}}{1+e^{\hat{\hat{\bm \beta}}' \hat{\bm X}_i}}\right) \hat{\bm X}_i     \\
      \hat{\hat \kappa} - \frac{1}{n}\sum_{i=1}^n \hat A_i \left( 1+ e^{h(\hat{\bm X}_i,  \hat Y_i)-\hat{\hat{\bm \beta}}'\hat{\bm X}_i}\right)     \\
      \hat{\hat \kappa }  \hat{\hat \mu}^{(h)} - \frac{1}{n}\sum_{i=1}^n \hat A_i \hat Y_i \left( 1+ e^{h(\hat{\bm X}_i,  \hat Y_i)-\hat{\hat{\bm \beta}}'\hat{\bm X}_i}\right)
    \end{array}
  \right) = 0.
\end{align}

Now, invoking the asymptotic theory of bootstrap for $Z$-estimators
\citep[Chapter 10]{bootz,korosok}, we can derive validity of the
bootstrap confidence intervals discussed in Section
\ref{sec:nonp-bootstr}. 
As a remark, the result of Wellner and Zhan (1996) extend to
nonparametric bootstrap of infinite dimensional $Z$-estimators, under
a collection of regularity conditions. Therefore, we can expect that
the bootstrap procedure introduced above to also work if the missing
probability is modeled non-parametrically, if the model satisfies
these regularity conditions. However, it is important to note that
bootstrap is generally not valid for general non-parametric models, as
observed by \citet{abadie2008failure}.

To this end, we need the following assumption.

\begin{assumption}\label{assum:moment} The parameter space $\Theta$ is compact and the true parameter $\bm \upsilon_0$ is in the interior of $\Theta$. Moreover, the joint distribution of $(Y, \bm X)$ satisfies:
  \begin{itemize}
  \item[(1)] $\E [Y^4]< \infty$.

  \item[(2)] $\mathrm{det}(\E \left( \frac{e^{\bm \beta_0' \bm X}}{(1+e^{\bm \beta_0' \bm X})^2} \bm X \bm X' \right)) >0$.

  \item[(3)]  For every compact subset $S\subset \R^d$, $\E\left[ \sup_{\bm \beta \in S} e^{\bm \beta' \bm X} \right] < \infty$.
    \end{itemize}
\end{assumption}

Note that all the assumptions are trivially satisfied if we assume that the supports of $\bm X$ and $Y$ are bounded and  $\E[\bm X \bm X']$ is positive definite. The assumptions allow for more general distributions, for example, distributions with $\E[e^{t ||\bm X||}]< \infty$, for all $t \in \R$. Under this assumption we can derive the asymptotic limiting distribution of the bootstrap estimates (recall  \eqref{eq:phi} and \eqref{eq:phihat}).

\begin{thm}\label{thm:normality} Suppose the joint distribution of $(Y, \bm X)$ satisfies Assumption \ref{assum:moment}. Then, for  $h \in \cH(\gamma)$ be fixed, under $\P_0$,
  $$\sqrt n\left(\hat{\bm \upsilon}- \bm \upsilon \right) \dto N(\bm 0, \dot \Phi_0^{-1} \Sigma \dot \Phi_0), \quad \text{and} \quad \sqrt n\left(\hat{\hat{\bm \upsilon}} - \bm \upsilon \right) \dto N(\bm 0, \dot \Phi_0^{-1} \Sigma \dot \Phi_0),$$
  where
  \begin{align}\label{eq:phi0}
    \dot \Phi_0 = \E \grad_{\bm \upsilon=\bm \upsilon_0}Q(\bm T|\bm \upsilon)=
    \left(
      \begin{array}{ccc}
        \bm 0 & \bm 0 & -\E \left( \frac{e^{\bm \beta_0' \bm X}}{(1+e^{\bm \beta_0' \bm X})^2}\right) \bm X \bm X'   \\
        0 & 1  &    \E A \bm X' e^{h(\bm X,  Y)-\bm \beta_0' \bm X}  \\
        \kappa^{(h)} & \mu^{(h)} &   \E A Y \bm X' e^{h(\bm X,  Y)-\bm \beta_0' \bm X}
      \end{array}
    \right),
  \end{align}
  and $\Sigma:=\E[Q(\bm T, \bm \upsilon_0)Q(\bm T, \bm \upsilon_0)']$.
\end{thm}

The limiting distribution of $\hat \mu^{(h)}$ (recall \eqref{eq:sipw} and \eqref{eq:phi}) and $\hat{\hat \mu}^{(h)}$ (defined through \eqref{eq:phihat}) is an immediate consequence of the above theorem:

\begin{cor}\label{cor:normality} Suppose the joint distribution of $(Y, \bm X)$ satisfies Assumption \ref{assum:moment}. Then, for  $h \in \cH(\gamma)$ be fixed, under $\P_0$,
  $$\sqrt n\left(\hat{\mu}^{(h)}-\mu^{(h)}\right) \dto N(0, (\sigma^{(h)})^2), \quad \text{and} \quad \sqrt n\left(\hat{\hat{\mu}}^{(h)}-\hat{\mu}^{(h)}\right) \dto N(0, (\sigma^{(h)})^2),$$
  where $(\sigma^{(h)})^2=(\dot \Phi_0^{-1} \Sigma \dot \Phi_0)_{11}$ is the first diagonal element of $\dot \Phi_0^{-1} \Sigma \dot \Phi_0$.
\end{cor}

The rest of the section is organized as follows: In Section \ref{sec:pfnormal} we prove Theorem \ref{thm:normality}, and in Section \ref{sec:pfvalidpbz} we complete the proof of Theorem \ref{thm:valid-pbz} using Corollary \ref{cor:normality}.

\subsection{Proof of Theorem \ref{thm:normality}}
\label{sec:pfnormal}

We begin by verifying that the matrices $\dot \Phi_0^{-1}$ and $\Sigma$ in Theorem \ref{thm:normality} are well-defined. To begin with, a direct computation gives 
\begin{align*}
  |\det(\dot \Phi_0)| & = |\det \left(
    \begin{array}{ccc}
      -\E \left( \frac{e^{\bm \beta_0' \bm X}}{(1+e^{\bm \beta_0' \bm X})^2}\right) \bm X \bm X' & \bm 0 & \bm 0   \\
      \E A \bm X' e^{h(\bm X,  Y)-\bm \beta_0' \bm X} & 0 & 1    \\
      \E A Y \bm X' e^{h(\bm X,  Y)-\bm \beta_0' \bm X} & \kappa^{(h)} & \mu^{(h)}
    \end{array}
  \right)|= \kappa^{(h)}
  |\det \E \left( \frac{e^{\bm \beta_0' \bm X}}{(1+e^{\bm \beta_0' \bm X})^2} \bm X \bm X' \right)| > 0,
\end{align*}
by Assumption \ref{assum:moment} (2). Therefore, $\dot \Phi_0$ is invertible. Also, note that $\Sigma < \infty$, which follows by direct multiplication and Assumption \ref{assum:moment} (1).

We can now proceed to prove Theorem \ref{thm:normality}. This will be done by invoking  \citep[Theorem 10.16]{korosok}, which gives conditions are asymptotic normality of bootstrapped $Z$-estimators. This entails verifying the following three conditions:
\begin{itemize}

\item[(A)] The class of functions $\{\bm t \rightarrow Q(\bm t|\bm \upsilon): \bm \upsilon \in \Theta\}$ is $\P_0$-Glivenko-Cantelli (proved in Section \ref{sec:pfA}).

\item[(B)] $||\Phi(\bm \upsilon)||_1$ is strictly positive outside every open neighborhood of $\bm \upsilon_0$ (proved in Section \ref{sec:pfB}).

\item[(C)] The class of functions $\{\bm t \rightarrow Q(\bm t|\bm \upsilon): \bm \upsilon \in \Theta\}$ is $\P_0$-Donsker and $\E[\left(Q(\bm T|\bm \upsilon_n)-Q(\bm T|\bm \upsilon_0)\right)^2] \rightarrow 0$, whenever $||\bm \upsilon_n-\bm \upsilon_0||_1 \rightarrow 0$ (proved in Section \ref{sec:pfC}).

\end{itemize}


\subsubsection{\textbf{Proof of} (A)}
\label{sec:pfA}

Define the envelope function $B(\bm t):=\sup_{\bm \upsilon \in \Theta}||Q(\bm t|\bm \upsilon)||_1$. Then using the compactness of $\Theta$, $||h||_\infty \leq \gamma$, and $|a| \leq 1$,
\begin{align}
  ||Q(\bm t|\bm \upsilon)||_1 & \leq ||Q_1(\bm t|\bm \upsilon)||_1+|Q_2(\bm t|\bm \upsilon)|+|Q_3(\bm t|\bm \upsilon)| \nonumber \\
  & \leq  ||\bm x||_1 + |y|+ e^{\gamma}e^{-\bm \beta' \bm x} (1+|y|) + M,
\end{align}
for some absolute constant $M$. Therefore, by Assumption \ref{assum:moment}, $\E B(\bm T) < \infty$, and by \citep[Lemma 6.1]{lecture_jw}, the class of functions $\{\bm t \rightarrow Q(\bm t|\bm \upsilon): \bm \upsilon \in \Theta\}$ is $\P_0$-Glivenko-Cantelli.

\subsubsection{\textbf{Proof of} (B)}
\label{sec:pfB}

To begin with note that by Assumption \ref{assum:moment} (2), $$\E \left( \frac{e^{\bm \beta' \bm X}}{(1+e^{\bm \beta' \bm X})} \bm X \right)=0,$$
has a unique root $\bm \beta=\bm \beta_0$, since its gradient is positive definite at $\bm \beta_0$, and non-negative definite everywhere.  Then, fixing $\varepsilon > 0$, we have
\begin{align}\label{eq:Phi1}
  ||\Phi(\bm \upsilon)||_1 \geq \left|\left|\E\left[ \frac{e^{\bm \beta'_0 \bm X}}{1+e^{\bm \beta'_0 \bm X}} \bm X- \frac{e^{\bm \beta' \bm X}}{1+e^{\bm \beta' \bm X}} \bm X\right] \right|\right|_1  >0,
\end{align}
whenever $||\bm \beta-\bm \beta_0||_1 > \frac{\varepsilon}{M}$, where $M$ is a constant to be chosen later.

Next, assume that $||\bm \beta-\bm \beta_0||_1 \leq \frac{\varepsilon}{M}$. This implies $||\bm \beta-\bm \beta_0||_\infty \leq \frac{\varepsilon}{M}$ and
\begin{align}\label{eq:Phi1leq}
  \left|\E\left[  A e^{h(\bm X, Y)- \bm \beta'_0 \bm X} - A e^{h(\bm X, Y)- \bm \beta' \bm X} \right]\right|& \leq \E\left[  \left|A e^{h(\bm X, Y)}\right| \cdot \left |e^{- \bm \beta'_0 \bm X} - e^{ -\bm \beta' \bm X} \right|\right] \nonumber \\
  & \leq e^\gamma \E \left |e^{- \bm \beta'_0 \bm X} - e^{- \bm \beta' \bm X} \right| \nonumber \\
  & \leq  e^\gamma \E \left|(\bm \beta-\bm \beta_0)' \bm X e^{-t_*} \right|  \nonumber \tag{for some $t_* \in [\bm \beta_0' \bm X, \bm \beta'\bm  X ]$} \\
  & \leq  e^\gamma  ||\bm \beta-\bm \beta_0||_\infty \E \left[ ||\bm X||_1e^{-t_*} \right] \nonumber  \\
    & \leq  e^\gamma  ||\bm \beta-\bm \beta_0||_\infty \left( \E [ ||\bm X||_1^2 \E [e^{-2t_*}]\right)^{\frac{1}{2}} \nonumber  \\
  & \leq   ||\bm \beta-\bm \beta_0||_\infty \left(e^{2\gamma}  \E [ ||\bm X||_1^2] \cdot  \E\left[\sup_{\bm \beta_*: ||\bm \beta_*-\bm \beta_0||_1 \leq \frac{\varepsilon}{M}} e^{-2\bm \beta_* \bm X}\right] \right)^{\frac{1}{2}}\nonumber  \\
  & \leq K_1(\gamma) \cdot \frac{ \varepsilon}{M} \leq \frac{\varepsilon}{64 K}.
\end{align}
by choosing $M \geq 64 K \cdot K_1(\gamma)$, where
$$K_1(\gamma)^2:=e^{2\gamma}  \E [ ||\bm X||_1^2] \cdot \E\left[\sup_{\bm \beta_*: ||\bm \beta_*-\bm \beta_0||_1 \leq \frac{\varepsilon}{M}} e^{-2\bm \beta_* \bm X}\right] < \infty,$$ by Assumption \ref{assum:moment},
and $K:=\sup_{\bm \upsilon \in \Theta} |\nu| \in (0, \infty)$ by  the compactness of $\Theta$.
Therefore, whenever $||\bm \beta-\bm \beta_0||_1 \leq \frac{\varepsilon}{M}$ and $|\kappa-\kappa^{(h)}| > \frac{\varepsilon}{4K}$,
\begin{align}\label{eq:Phi2}
  ||\Phi(\bm \upsilon)||_1 \geq \left|\kappa-\kappa^{(h)}+\E\left[  A e^{h(\bm X, Y)- \bm \beta'_0 \bm X} - A e^{h(\bm X, Y)- \bm \beta' \bm X} \right] \right| >0.
\end{align}

Finally, assume that $||\bm \beta-\bm \beta_0||_1 \leq \frac{\varepsilon}{M}$ and $|\kappa-\kappa^{(h)}| \leq \frac{\varepsilon}{4K}$, but  $|\nu-\mu^{(h)}| > \frac{\varepsilon}{2\kappa^{(h)}}$. Then, as in \eqref{eq:Phi2},
\begin{align}\label{eq:Phi2leq}
  \left|\E\left[  A Ye^{h(\bm X, Y)- \bm \beta'_0 \bm X} - A Y e^{h(\bm X, Y)- \bm \beta' \bm X} \right]\right| & \leq K_2(\gamma) \cdot \frac{ \varepsilon}{M} \leq \frac{\varepsilon}{64 K},
\end{align}
by choosing $M \geq 64 K \cdot K_2(\gamma)$, where
$$K_2(\gamma)^2:=e^{2\gamma}  \E [ ||Y\bm X||_1^2] \cdot \E\left[\sup_{\bm \beta_*: ||\bm \beta_*-\bm \beta_0||_1 \leq \frac{\varepsilon}{M}} e^{-2\bm \beta_* \bm X}\right] < \infty,$$ by Assumption \ref{assum:moment}. Using \eqref{eq:Phi2leq} and  $|\kappa^{(h)} \nu-\kappa \nu| \leq \frac{\varepsilon}{4}$ now gives
\begin{align}\label{eq:Phi3}
  ||\Phi(\bm \upsilon)||_1=\left|\kappa \nu-\kappa^{(h)} \mu^{(h)}+\E\left[  A Y e^{h(\bm X, Y)- \bm \beta'_0 \bm X} - A Ye^{h(\bm X, Y)- \bm \beta' \bm X} \right] \right| >0.
\end{align}

Combining \eqref{eq:Phi1}, \eqref{eq:Phi2}, and \eqref{eq:Phi3}, we get, for all $\delta >0$, $\inf\{||\Phi(\bm \upsilon)||^2: ||\bm \upsilon-\bm \upsilon_0||_1 > \delta \} >0$, as required.


%

\subsubsection{\textbf{Proof of} (C)}
\label{sec:pfC}


We begin by showing the class of functions $\{\bm t \rightarrow Q(\bm t|\bm \upsilon): \bm \upsilon \in \Theta\}$ is $\P_0$-Donsker. To this end, let $\bm \upsilon_1=(\nu_1, \kappa_1, \bm \beta_1')'$ and $\bm \upsilon_2=(\nu_2, \kappa_2, \bm \beta_2')'$ be two points in the parameter space $\Theta$, and recall the definition of $Q(\cdot|\bm \upsilon)$ from \eqref{eq:Q}. Then, by the mean-value theorem, there exists $t_* \in \R$ such that $t_* \in [\bm \beta_1' \bm x, \bm \beta_2'\bm  x ]$ such that\footnote{For $x, y \in \R$, $x \lesssim y$ means $x \leq C y$, for some constant $C>0$. We will use subscripts $\lesssim_\square$,  to denote that the constant may depend on the subscripted parameters.}
\begin{align}\label{eq:diff1}
  ||Q_1(\bm t|\bm \upsilon_2)-Q_1(\bm t|\bm \upsilon_1)||_1  \leq  \left|\left|\left(\frac{e^{\bm \beta_2' \bm x}}{1+e^{t_*}} - \frac{e^{\bm \beta_1' \bm x}}{1+e^{\bm \beta_1' \bm x}} \right) \bm x \right|\right|_1 & \leq \left|\left|\left(\frac{e^{t_*}}{(1+e^{t_*})^2}  \right) (\bm \beta_2-\bm \beta_1)' \bm x \bm x \right|\right|_1 \nonumber \\
  & \leq |(\bm \beta_2-\bm \beta_1)' \bm x| \left|\left| \bm x \right|\right|_1 \nonumber \\
  & \leq ||\bm \beta_2-\bm \beta_1'||_2 ||\bm x||_2||\bm x ||_1 \nonumber \\
  & \lesssim_d M_1(\bm x) ||\bm \beta_2-\bm \beta_1'||_1,
\end{align}
where $M_1(\bm x) =||\bm x ||_1^2$. Next, observe that
\begin{align}
  |Q_2(\bm t|\bm \upsilon_2)-Q_2(\bm t|\bm \upsilon_1)| & \leq |\kappa_2-\kappa_1|+ e^{\gamma}|e^{-\bm \beta_2' \bm x}-e^{-\bm \beta_1' \bm x}| \nonumber \\
  & \lesssim_\gamma |\kappa_2-\kappa_1|+ \left|(\bm \beta_2-\bm \beta_1)' \bm x e^{-t_*} \right|  \nonumber \tag{for some $t_* \in [\bm \beta_1' \bm x, \bm \beta_2'\bm  x ]$} \\
  & \lesssim_\gamma |\kappa_2-\kappa_1|+ ||\bm \beta_2-\bm \beta_1||_2 \cdot ||\bm x||_2 \sup_{\bm \beta \in \Theta_0} e^{-\bm \beta' \bm x} \nonumber \\
  & \lesssim_{\gamma, d} M_2(\bm x)\left( |\kappa_2-\kappa_1|+ ||\bm \beta_2-\bm \beta_1||_1  \right),   \label{eq:diff2}
\end{align}
where $M_2(\bm x) =1+||\bm x||_1 \sup_{\bm \beta \in \Theta_0} e^{-\bm \beta' \bm x}$ and $\Theta_0$ is the projection of the parameter space to the last $d$ coordinates. Finally,
\begin{align}\label{eq:diff3}
  |Q_3(\bm t|\bm \upsilon_2)-Q_2(\bm t|\bm \upsilon_1)| & \leq |\kappa_2\nu_2-\kappa_1\nu_1|+ e^{\gamma}|y e^{-\bm \beta_2' \bm x}- y e^{-\bm \beta_1' \bm x}| \nonumber \\
  & \lesssim_\gamma |\kappa_2-\kappa_1|+ |\nu_2-\nu_1|+||\bm \beta_2-\bm \beta_1'||_2 \cdot ||y \bm x||_2 \sup_{\bm \beta \in \Theta_0} e^{-\bm \beta' \bm x}  \nonumber \\
  & \lesssim_{\gamma, d} M_3(\bm x, y)\left( |\kappa_2-\kappa_1|+ |\nu_2-\nu_1|+ ||\bm \beta_2-\bm \beta_1||_1  \right)  \nonumber
\end{align}
where $M_3(\bm x, y) =1+||y \bm x||_1 \sup_{\bm \beta \in \Theta_0} e^{-\bm \beta' \bm x}$.

Therefore, defining $M(\bm x, y)=M_1(\bm x)+M_2(\bm x)+M_3(\bm x, y)$ and combining \eqref{eq:diff1},  \eqref{eq:diff2}, and \eqref{eq:diff3} gives
\begin{align}
  ||Q(\bm t|\bm \upsilon_2)-Q(\bm t|\bm \upsilon_1)||_1  = \sum_{b=1}^3 ||Q_b(\bm t|\bm \upsilon_2)-Q_b(\bm t|\bm \upsilon_1)||_1 \lesssim_{\gamma, d} M(\bm x, y) ||\bm \upsilon_2-\bm \upsilon_1||_1
\end{align}
Since $\E[M(\bm X, Y)^2] < \infty$, by Assumption \ref{assum:moment}, this implies that the class $\{\bm t \rightarrow Q(\bm t|\bm \upsilon): \bm \upsilon \in \Theta\}$ is $\P$-Donsker.


The inequalities in \eqref{eq:diff1},  \eqref{eq:diff2}, and
\eqref{eq:diff3}, combined with Assumption \ref{assum:moment} also
implies that $$\E[\left(Q(\bm T|\bm \upsilon_n)-Q(\bm T|\bm
  \upsilon_0)\right)^2] \rightarrow 0$$
holds coordinate-wise, whenever $||\bm \upsilon_n-\bm \upsilon_0||_1 \rightarrow 0$.

\subsection{Completing the Proof of Theorem \ref{thm:valid-pbz}}
\label{sec:pfvalidpbz}

To begin with, note that the $\alpha$-th bootstrap quantile $Q_\alpha(\hat{\hat{\mu}}^{(h)})=\hat \mu^{(h)}+\frac{\hat z_{\alpha, n}^{(h)}}{\sqrt n}$, where $\hat z_{\alpha, n}^{(h)}:=\inf\{t: \hat {\mathbb P}_n(\sqrt n(\hat{\hat{\mu}}^{(h)}-\hat{\mu}^{(h)}) \leq t) \geq \alpha\}$. By Corollary \ref{cor:normality}, as $n \rightarrow \infty$, $\hat z_{\alpha, n}^{(h)} = z_{\alpha}^{(h)}+O_P(1/\sqrt n)$, where $z_{\alpha}^{(h)}$ is $\alpha$-th quantile of $N(0, (\sigma^{(h)})^2)$. Then, recalling \eqref{eq:lbh}, as $n \rightarrow \infty$,
\begin{align*}
  \P_0\left(\mu^{(h)} < L^{(h)}_B \right)=\P_0\left(\sqrt n\left(\hat \mu^{(h)} -\mu^{(h)} \right) \leq z_{\frac{\alpha}{2}, n}^{(h)} \right) \rightarrow \frac{\alpha}{2},
\end{align*}
by Corollary \ref{cor:normality}. The limit of  $\P_0\left(\mu^{(h)} > U^{(h)}_B \right)$ follows similarly.

%

\section{Proof of Corollary \ref{cor:valid-pb-obs}}

The proof of Corollary \ref{cor:valid-pb-obs} is similar to the proof of Theorem \ref{thm:valid-pb}. We begin by defining
$$\kappa_0^{(h)}:=\E \left[\frac{1-A}{e^{(h)}_0(\bm X,Y)}\right], \quad \kappa_1^{(h)}:=\E \left[\frac{A}{e^{(h)}_1(\bm X, Y)}\right].$$
We now set up the SIPW estimate of the ATE as a $Z$-estimation problem. To this end, given a vector $\bm \upsilon=(\nu_0, \nu_1, \kappa_0^{(h)}, \kappa_1^{(h)} ,\bm \beta')' \in \Theta \subset \R \times \R \times \R_+\times \R_+ \times \R^d$, where $\Theta$ is the parameter space, define the function $Q: \{0, 1\} \times \R \times \R^d \rightarrow \R^{d+4}$ as follows: For $\bm t=(a, \bm x', y)' \in \{0, 1\} \times \R \times \R^d$,

\begin{align}\label{eq:Qobs}
  Q(\bm t|\bm \upsilon)= \left(
    \begin{array}{c}
      Q_1(\bm t|\bm \upsilon)   \\
      Q_2(\bm t|\bm \upsilon)    \\
      Q_3(\bm t|\bm \upsilon) \\
      Q_4(\bm t|\bm \upsilon) \\
      Q_5(\bm t|\bm \upsilon)
    \end{array}
  \right):=
  \left(
    \begin{array}{c}
      \left(a-\frac{e^{\bm \beta' \bm x}}{1+e^{\bm \beta' \bm x}}\right) \bm x     \\
      \kappa_1^{(h)} - a \left( 1+ e^{h_1(\bm x,  y)-\bm \beta' \bm x}\right)     \\
      \kappa_1^{(h)}  \nu_1 - a y \left( 1+ e^{h_1(\bm x,  y)-\bm \beta' \bm x}\right) \\
      \kappa_0^{(h)} - (1-a) \left( 1+ e^{-h_0(\bm x,  y)+\bm \beta' \bm x}\right)     \\
      \kappa_0^{(h)}  \nu_0 - (1-a) y \left( 1+ e^{-h_0(\bm x,  y)+\bm \beta' \bm x}\right)
    \end{array}
  \right).
\end{align}
Next, define $\Phi(\bm \upsilon)=\int Q(\bm t|\bm \upsilon) \mathrm d\P_0(\bm t)$, where  $\bm T=(A, \bm X', Y)' \sim \P_0$, the true distribution generating the data. Note that $\Phi(\bm \upsilon_0)=0$, where  $\bm \upsilon_0=(\mu^{(h_0)}(0), \mu^{(h_1)}(1), \kappa_0^{(h)}, \kappa_1^{(h)}, \bm \beta_0')$ is the true parameter value. Then, as in \eqref{eq:phi}, $Z$-estimates $\hat {\bm \upsilon}=(\hat \mu^{(h_0)}(0), \hat \mu^{(h_1)}(1), \hat \kappa^{(h_0)}_0, \hat \kappa^{(h_1)}_1, \hat{\bm \beta}')' $ are obtained by solving the equations
\begin{align*}
  \Phi_n(\hat {\bm \upsilon}):=\frac{1}{n}\sum_{i=1}^n Q(\bm T_i|\hat {\bm \upsilon})= 0.
\end{align*}
It is easy to see that the $Z$ estimate $\hat \mu^{(h_1)}(1)-\hat \mu^{(h_0)}(0)$ is exactly the SIPW estimate for $\Delta^{(h_0, h_1)}$ and $\hat {\bm \beta}$ is the MLE of $\bm \beta$ for the logistic regression model. Moreover, as in \eqref{eq:phihat}, the bootstrap $Z$-estimates $\hat{\hat {\bm \upsilon}}$ are obtained from the equations:
\begin{align*}
  \hat \Phi_n(\hat{\hat {\bm \upsilon}}):=\frac{1}{n}\sum_{i=1}^n Q(\hat{\bm T_i}|\hat{\hat {\bm \upsilon}})= 0.
\end{align*}

Then, as in Theorem \ref{thm:normality}, the limiting joint normality
of $(\hat \mu^{(h_0)}(0), \hat \mu^{(h_1)}(1))$, and hence the
asymptotic normality of $\hat \Delta^{(h_0, h_1)}=\hat
\mu^{(h_0)}(0)-\hat \mu^{(h_1)}(1)$, can be derived. This would imply
the asymptotic validity of the confidence interval
\eqref{eq:sen_interval_obs}, as in Section \ref{sec:pfvalidpbz},
completing the proof of Corollary \ref{cor:valid-pb-obs}. Details are
omitted.


\end{document}